\documentclass[12pt,epsf]{article}
\usepackage{verbatim}
\usepackage{anyfontsize}
\usepackage{dsfont}
\usepackage{tikz}
\usepackage{sidecap}
\sidecaptionvpos{figure}{t}
\usepackage{subfigure}
\usepackage{enumitem}
\usepackage[english]{babel}
\usepackage{enumitem}  

\usepackage{amssymb,amsmath}
\usepackage{graphicx, xcolor, varwidth}
\usepackage{setspace}
\usepackage[permil]{overpic}
\usepackage{cite}
\usepackage{mathtools}

\DeclareSymbolFont{matha}{OML}{txmi}{m}{it}
\DeclareMathSymbol{v}{\mathord}{matha}{118}

\colorlet{darkblue}{blue!70!black}
\colorlet{darkgreen}{green!70!black}

\usepackage[colorlinks=true,urlcolor=darkblue,linktocpage=true,linkcolor=darkblue,citecolor=darkblue]{hyperref}

\numberwithin{equation}{section}
\DeclareMathSymbol{v}{\mathord}{matha}{118}
\newcommand{\be}{\begin{equation}}
\newcommand{\ee}{\end{equation}}
\newcommand{\bea}{\begin{eqnarray}}
\newcommand{\eea}{\end{eqnarray}}
\newcommand{\bear}{\begin{eqnarray}}
\newcommand{\eear}{\end{eqnarray}}
\newcommand{\beas}{\begin{eqnarray*}}

\newcommand{\eeas}{\end{eqnarray*}}
\newcommand{\ba}{\begin{array}}
\newcommand{\ea}{\end{array}}

\def\ba#1\ea{\begin{align}#1\end{align}}
\def\bs#1\es{\begin{split}#1\end{split}}

\newcommand{\blockS}[5]{
\begin{tikzpicture}[baseline=-3pt]
\coordinate (v1) at (-1,-0.5) {} {};
\coordinate (v2) at (-0.5,0) {} {};
\coordinate (v3) at (-1,0.5) {} {};
\coordinate (v4) at (0.5,0) {} {};
\coordinate (v5) at (1,0.5) {} {};
\coordinate (v6) at (1,-0.5) {} {};
\begin{scope}[very thick]
\draw  (v1) node[left] {$#2$}--(v2);
\draw  (v3) node[left] {$#1$}-- (v2);
\draw  (v4) -- (v2) node[midway, above] {$#5$};
\draw  (v5) node[right] {$#3$}-- (v4);
\draw  (v6) node[right] {$#4$}-- (v4);
\end{scope}
\end{tikzpicture}
}

\newcommand{\blockTT}[5]{
\begin{tikzpicture}[baseline=-3pt]
\coordinate (v1) at (-1.5,-0.5) {} {};
\coordinate (v2) at (-1,0) {} {};
\coordinate (v3) at (-1.5,0.5) {} {};
\coordinate (v4) at (1,0) {} {};
\coordinate (v5) at (1.5,0.5) {} {};
\coordinate (v6) at (1.5,-0.5) {} {};
\begin{scope}[very thick]
\draw  (v1) node[left] {$#2$}--(v2);
\draw  (v3) node[left] {$#1$}-- (v2);
\draw  (v4) -- (v2) node[midway, above] {$#5$};
\draw  (v5) node[right] {$#3$}-- (v4);
\draw  (v6) node[right] {$#4$}-- (v4);
\end{scope}
\end{tikzpicture}
}

\newcommand{\blockT}[5]{
\begin{tikzpicture}[baseline=-3pt]
\coordinate (v1) at (-1.3,-0.5) {} {};
\coordinate (v2) at (-0.8,0) {} {};
\coordinate (v3) at (-1.3,0.5) {} {};
\coordinate (v4) at (0.8,0) {} {};
\coordinate (v5) at (1.3,0.5) {} {};
\coordinate (v6) at (1.3,-0.5) {} {};
\begin{scope}[very thick]
\draw  (v1) node[left] {$#2$}--(v2);
\draw  (v3) node[left] {$#1$}-- (v2);
\draw  (v4) -- (v2) node[midway, above] {$#5$};
\draw  (v5) node[right] {$#3$}-- (v4);
\draw  (v6) node[right] {$#4$}-- (v4);
\end{scope}
\end{tikzpicture}
}

\newcommand{\blocklarge}[5]{
\begin{tikzpicture}[baseline=-3pt]
\coordinate (v1) at (-2.0,-0.5) {} {};
\coordinate (v2) at (-1.5,0) {} {};
\coordinate (v3) at (-2,0.5) {} {};
\coordinate (v4) at (1.5,0) {} {};
\coordinate (v5) at (2,0.5) {} {};
\coordinate (v6) at (2,-0.5) {} {};
\begin{scope}[very thick]
\draw  (v1) node[left] {$#2$}--(v2);
\draw  (v3) node[left] {$#1$}-- (v2);
\draw  (v4) -- (v2) node[midway, above] {$#5$};
\draw  (v5) node[right] {$#3$}-- (v4);
\draw  (v6) node[right] {$#4$}-- (v4);
\end{scope}
\end{tikzpicture}
}

\newcommand{\tr}{\operatorname{tr}}
\newcommand{\pd}[2][1]{\ifnum#1=1 \frac{\partial}{\partial {#2}} \else
  \frac{\partial^#1}{\partial {#2}^{#1}}\fi}
\newcommand{\dpd}[2][1]{\ifnum#1=1 \dfrac{\partial}{\partial {#2}} \else
  \frac{\partial^#1}{\partial {#2}^{#1}}\fi}
\newcommand{\td}[2][1]{\ifnum#1=1 \frac{d}{d{#2}} \else
  \frac{d^#1}{d{#2}^{#1}}\fi}

\newcommand{\x}{\xi}

\newcommand{\nbox}{{\,\lower0.9pt\vbox{\hrule \hbox{\vrule height 0.2 cm \hskip 0.19 cm \vrule height 0.2 cm}\hrule}\,}}

\def\O{{\cal O}}

\newcommand{\cG}{{\cal G}}
\newcommand{\CO}{\mathcal{O}}

\newcommand{\bz}{\bar{z}}

\newcommand{\N}{{\cal N}}


\begin{document}
\begin{spacing}{1.3}
\begin{titlepage}

\begin{center}
{\Large \bf 
A Species or Weak-Gravity Bound for
\\    Large $N$ Gauge Theories Coupled to Gravity
}

\vspace*{6mm}

Jared Kaplan and Sandipan Kundu

\vspace*{6mm}

\textit{Department of Physics and Astronomy, Johns Hopkins University,
Baltimore, Maryland, USA\\}

\vspace{6mm}

{\tt \small jaredk@jhu.edu, kundu@jhu.edu}

\vspace*{6mm}
\end{center}

\begin{abstract}

Causality  constrains the gravitational interactions of massive higher spin particles in both AdS and flat spacetime.  We explore the extent to which these constraints apply to composite particles, explaining why they do not rule out macroscopic objects or hydrogen atoms.  However, we find that they do apply to glueballs and mesons in confining large $N$ gauge theories.  Assuming such theories contain massive bound states of general spin, we find parametric bounds in $(3+1)$ spacetime dimensions of the form $N\lesssim \frac{M_{Pl}}{\Lambda_{\text{QCD}}}$ relating $N$, the QCD scale, and the Planck scale.  We also argue that a stronger bound replacing $\Lambda_{\text{QCD}}$ with the UV cut-off scale may be derived from eikonal scattering in flat spacetime. 

\end{abstract}

\end{titlepage}
\end{spacing}

\vskip 1cm
\setcounter{tocdepth}{2}  
\tableofcontents

\begin{spacing}{1.3}

\section{Introduction}

Constraints on particle interactions seem to grow more and more stringent with increasing spin.  Most famously, massless particles of spin $1$ and $2$ must couple to charge or energy-momentum if they are to mediate long-range forces, while higher spin particles must decouple at long distances \cite{Weinberg:1965nx, Benincasa:2007xk}.  

More general constraints on massive higher spin particles can be derived from causality  \cite{Afkhami-Jeddi:2018apj}, ruling out some effective field theories and circumscribing the allowed UV completions of others.   Consistent quantum field theories (QFTs) may also be constrained by rather different bounds relating to the total number of particle species \cite{Bekenstein:1980jp, Casini:2008cr, Susskind:1994sm, Bousso:1999xy, Veneziano:2001ah, Dvali:2007hz}  or the relative strength of the gravitational force \cite{ArkaniHamed:2006dz}.  Our  goal is to demonstrate a connection between these constraints in the case of confining large $N$ gauge theories.  In short, bounds on the interactions of long-lived higher spin hadrons in large $N$ gauge theories imply a relation between their effective coupling $\frac{1}{N^2}$ and the gravitational coupling $G_N$.

In this paper, we explore when large $N$ gauge theories can be coupled to gravity in a way that preserves unitarity, causality, and Lorentz invariance. In general, it is easier to show that a QFT is not well behaved than to show it is well behaved -- a single inconsistency is sufficient to conclude that the QFT is in the swampland. We argue that large $N$ gauge theories such as quantum chromodynamics (QCD) when coupled to Einstein gravity can lead to such inconsistencies. In particular, in $(3+1)$-dimensions we conclude that  if a large $N$ gauge theory (i) is a confining theory and (ii) contains glueballs and mesons of spin $J>2$, it
violates causality unless 
\be\label{bound}
N\lesssim \frac{M_{Pl}}{\Lambda_{\text{QCD}}}\ ,
\ee
where, $\Lambda_{\text{QCD}}$ is the confinement scale and $M_{Pl}$ is the Planck scale.  This bound also implies that the gauge forces between the hadrons must be stronger than the gravitational forces between these particles, a result reminiscent of the weak-gravity conjecture \cite{ArkaniHamed:2006dz}.

Much of our discussion will concern the distinction between fundamental and composite particles, as our goal is to establish when the bounds of  \cite{Afkhami-Jeddi:2018apj} apply to composites.  Composite particles may differ  due to their finite size and substructure, which leads to the breakdown of effective field theory descriptions at distances of order their size.  Relatedly, high-energy scattering of  composites may be dominantly inelastic, so that composites  shatter when struck hard.  We find that this last issue  plays a leading role, making causality bounds inapplicable for macroscopic objects and weakly bound states such as hydrogen atoms.

We will examine causality bounds from both AdS/CFT, with a large $N$ gauge theory in the bulk of AdS, and from flat space scattering.   The CFT analysis of sections \ref{sec:ConstraintsinAdS} and \ref{sec:AdSAnalysis} is  simpler and clearer, as there are unambiguous causality bounds on CFT correlators \cite{Hartman:2015lfa,Hartman:2016lgu,Afkhami-Jeddi:2016ntf}, and the correlators themselves can be explicitly decomposed in conformal blocks.  Thus in the AdS/CFT analysis we can directly point to  the conformal blocks that alleviate causality bounds on gravitational scattering for weakly bound composites.   Free two-particle states in AdS with angular momentum $J\geq 2$ provide an illustrative extreme example of a weakly bound composite state where we can see  explicitly why causality bounds do not apply.  In addition, CFT analysis of high energy gravitational scattering of weakly bound composites also demonstrates that these particles do not have hard centers. 

In contrast, the eikonal scattering analysis of section \ref{sec:ScatteringAnalysis} requires some subtle tricks \cite{Camanho:2014apa}, such as the use of Bose-enhancement of the initial states, in order to derive a general causality bound.  However, the flat space thought experiment naturally imposes a stronger constraint replacing $\Lambda_{\text{QCD}}$ of bound (\ref{bound}) with the UV cut-off scale of the combined\footnote{Here the cutoff is the scale at which  massive particles or strings outside the gauge sector contribute to the scattering amplitudes. Also note that $M_{Pl}$ in both (\ref{bound}) and (\ref{bound1}) is the physical Planck scale.} gauge and gravity theory, so that
\be\label{bound1}
\Lambda_{\text{UV}} \lesssim \frac{M_{Pl}}{N}\ . 
\ee 
Furthermore, we examine how mixings and inelastic scattering affect the argument, and conclude that it cannot be applied to large objects like Kerr black holes or weakly bound composites.  The eikonal scattering argument is not merely the flat space limit of the AdS/CFT analysis.

The outline of this paper is as follows.  In section \ref{sec:BriefReviews} we provide abbreviated reviews of large $N$ scalings and the causality bounds of \cite{Hartman:2015lfa, Afkhami-Jeddi:2018apj,Hartman:2016lgu,Afkhami-Jeddi:2016ntf,Afkhami-Jeddi:2017rmx}.  In section \ref{sec:ConstraintsinAdS} we critically examine the AdS/CFT causality constraints to understand if and when they should apply to composite particles.  We apply this analysis to confining large $N$ gauge theories in AdS in section \ref{sec:AdSAnalysis}.  In section \ref{sec:ScatteringAnalysis} we provide an independent analysis based on eikonal scattering in flat space.  We review a somewhat hand-waving argument for a similar species bound in section \ref{sec:SpeciesBoundEntropy}, and we summarize our conclusions in section \ref{sec:Discussion}.

\section{Brief Reviews}
\label{sec:BriefReviews}

In this section we provide a quick summary of the large $N$ limit of gauge theories, CFT causality constraints, and specific constraints on higher spin particles interacting with scalars via gravity.  

\subsection{Large $N$ Expansion of Gauge Theories}

Let's briefly review\footnote{For a more extensive classic review see \cite{Witten:1979kh}. } expectations for the spectrum and interactions of large $N$  gauge theories in $(3+1)$ spacetime dimensions. These theories are characterized by  a confinement scale $\Lambda_{\text{QCD}}$ where the 't Hooft coupling $\lambda$ becomes strong.  This determines the characteristic physical size of  hadrons bound together by the confining force.  It also sets the mass scale of generic mesons, with exceptions including light pseudo-goldstone bosons and mesons formed from heavy quarks.  

In the large $N$ limit, only the subset of planar Feynman diagrams  survive, leading to major simplifications.  In particular, in the exact $N=\infty$ limit the mesons and glueballs behave as stable particles that are free and non-interacting \cite{tHooft:1973alw, tHooft:1974pnl,Witten:1979kh}.   This property of confining large $N$ gauge theories will allow us to study the extension of the bound of \cite{Afkhami-Jeddi:2018apj} to higher spin mesons and glueballs.  

In the large $N$ limit, masses of mesons and glueballs scale as $m_\pi, m_G \sim \O(N^0)$. Meson decay rates are of order $\O(1/\sqrt{N})$ and hence the lifetime of a meson is rather long $\sim \O(N)$. Glueballs are even more stable with typical lifetime of order $\O(N^2)$. Mixing of mesons with glueballs are also suppressed by the factor $\O(1/\sqrt{N})$.  In fact, one way to distinguish mesons ($\pi$) and glueballs ($G$) is via the scaling of their large $N$ couplings, which we list below  \cite{tHooft:1973alw, tHooft:1974pnl,Witten:1979kh}\begin{align}
& \langle \pi \pi \pi \rangle \sim \frac{1}{\sqrt{N}}\ , \qquad \langle G G G \rangle \sim \frac{1}{N}\ , \nonumber \\
& \langle \pi \pi G \rangle \sim \frac{1}{N}\ , \qquad \langle \pi G G  \rangle \sim \frac{1}{N^{3/2}}\ ,\nonumber\\ 
& \langle \pi \pi \pi \pi \rangle \sim \frac{1}{N}\ , \qquad \langle GGGG\rangle \sim \frac{1}{N^2}\ . 
\end{align}
These results describe the scaling of the correlators or scattering amplitudes, assuming that amplitudes for free propagation $\langle \pi \pi \rangle \sim \langle G G \rangle \sim N^{0}$ are normalized so that they are independent of $N$.  Baryons  may also be present, but their mass will scale as $\sim N \Lambda_{\text{QCD}}$, and so at large $N$ they will be very heavy. 

Later we will be interested in coupling $G$ and $\pi$ to gravity, and so we need to know the scaling of matrix elements of these particles with $T_{\mu \nu}$, the stress-energy tensor operator.  It has a natural normalization, determined by the fact that integrals of $T_{\mu \nu}$ generate the Poincar\'e group.  This means that, for example
\begin{align}
\label{eq:GGTScaling}
& \langle \pi \pi T_{\mu \nu} \rangle \sim 1\ , \qquad \langle G G  T_{\mu \nu} \rangle \sim 1\ , \nonumber \\
& \langle \pi \pi \pi T_{\mu \nu} \rangle \sim \frac{1}{\sqrt{N}}\ , \qquad \langle GGG T_{\mu \nu} \rangle \sim \frac{1}{N}\ .
\end{align}
When we couple to gravity, this produces the usual expectations for scaling with $G_N$ and $N$, so in particular 2-to-2 scattering via graviton exchange is proportional to $G_N$ but has no $N$-dependence.  Also note that off-diagonal matrix elements  $\langle G G' T_{\mu \nu} \rangle$ may be present (particularly for higher spin particles) without any $1/N$ suppression.

The correlators and scattering amplitudes of hadrons will be weakly coupled, but these composite particles may not be described\footnote{In the standard model, an EFT description is available because pions and other mesons are pseudo-goldstone bosons, and are therefore parametrically lighter than $\Lambda_{\text{QCD}}$.} by a convenient effective field theory.  This is due to the fact that both the mass of these states and the putative cutoff will be of order $\Lambda_{\text{QCD}}$.  Nevertheless, we can estimate the magnitude and rough behavior of interactions using symmetry, unitarity, and large $N$ scaling.  

Although we do not have a rigorous proof, we expect that large $N$ gauge theories contain meson and glueball bound states of general spin.  On physical grounds, we would expect that it's possible to construct color singlet states with high angular momentum by `spinning' quarks and gluons.  More formally, it is easy to construct gauge invariant local operators such as $\tr [ F_{\mu_1 \mu_2} \cdots F_{\mu_{\ell -1} \mu_{J}} ]$ with arbitrary spin, and when acting on the vacuum these should create high-spin hadrons.  In principle hadrons with spin $J \geq 2$ could have large masses, but we do not expect their masses to scale with $N$.  In what follows we will  assume that spin $J$ hadrons with mass of parametric order $\Lambda_{\text{QCD}}$ exist in the spectrum.

\subsection{Summary of CFT Causality Constraints}
\label{sec:SummaryCFTCausality}

Let us assume that our CFT$_{d \geq 3}$ includes a higher  spin ($J \geq 2$) primary operator $\cG_J$ and a scalar primary $\CO$.   

\subsubsection*{Four-Point Function}

\begin{figure}[h]
\begin{center}
\begin{tikzpicture}[baseline=-3pt, scale=1.35]

\begin{scope}[very thick,shift={(4,0)}]
\coordinate (v1) at (-1.5,-1.5) {};
\coordinate(v2) at (1.5,1.5) {};
\coordinate (v3) at (1.5,-1.5) {};
\coordinate(v4) at (-1.5,1.5) {};

\draw[thin,-latex]  (v1) -- (v2)node[left]{$v$};
\draw[thin,-latex]  (v3) -- (v4)node[right]{$u$};
\draw[ultra thick,red]  (1.4,-1.4) -- (-0.2,0.2);
\draw[ultra thick,red] (0,0) --  (-1.35,1.35);
\draw  (0,3) -- (3,0);
\draw  (0,3) -- (-3,0);
\draw  (0,-3) -- (3,0);
\draw  (0,-3) -- (-3,0);
\draw(-2,-0.3)node[above]{ $\int \overline{\cG_J}$};
\draw(1.9,-0.3)node[above]{ $\int \cG_J$};
\filldraw[blue]  (-1.5,0) circle (5 pt);
\filldraw[blue]  (1.5,0) circle (5pt);
\coordinate(v5) at (0,0) {};
\def \fac {.6};
\draw[
	scale=.5,samples=50,thick,blue,-latex,domain=2.5:4.6,variable=\y
	] 
		plot ({-\fac/(\y)-\fac*\y},{-\fac/(\y)+\fac*\y});
\draw[
	scale=.5,samples=50,thick,blue,-latex,domain=2.5:4.6,variable=\y,
	]
	plot ({\fac/(\y)+\fac*\y},{\fac/(\y)-\fac*\y});
\filldraw[black]  (-0.88,0.65) circle (1 pt);
\filldraw[black]  (0.88,-0.65) circle (1pt);
\draw(-0.88,0.65)node[left]{ $\O(u,v)$};
\draw(0.88,-0.65)node[right]{ $\O(-u,-v)$};

\end{scope}

\end{tikzpicture}
\end{center}
\caption{\label{config} \small Example of a Rindler symmetric four-point function. The operators $\cG_J$'s can be smeared over some regions in a Rindler reflection symmetric way as well. Note that this is just a schematic representation of the actual four-point function that we will use to derive the bound. In the actual correlator $F$, the operators $\cG_J$'s are smeared around an imaginary time value.}
\end{figure}
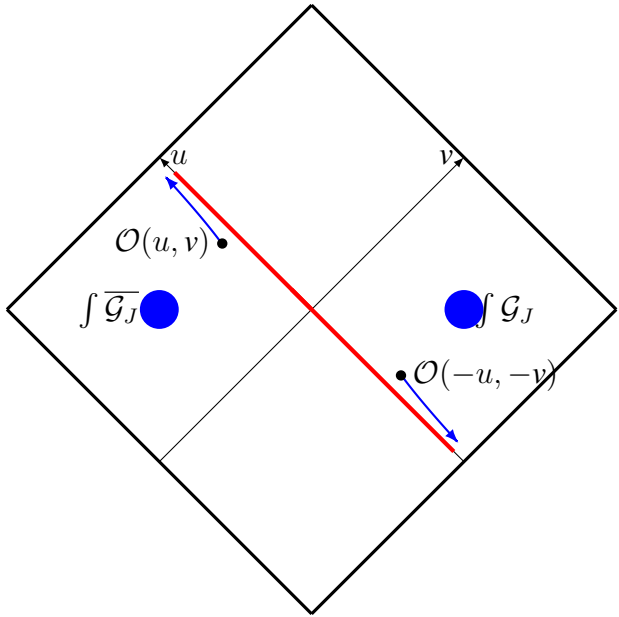

We will begin  with a normalized Rindler reflection symmetric\footnote{In CFT$_3$ coordinates of points are given by: $(t,y^1,y^2)$. Null coordinates $u$ and $v$ are defined as follows: $u=t-y^1$, $v=t+y^1$. For simplicity, whenever some coordinates are set to zero we will omit them.} Lorentzian four-point function
\be\label{maing}
F = \frac{\langle \overline{\varepsilon.\cG_J(B)} \CO(u, v) \CO( -u, - v) \varepsilon.\cG_J(B)\rangle}{\langle \overline{\varepsilon.\cG_J(B)}\varepsilon.\cG_J(B)\rangle \langle  \CO (u, v) \CO( -u, - v)\rangle} \ ,
\ee
as shown in figure \ref{config}, where our abbreviated notation implies
\be\label{ob}
\varepsilon.\cG_J(B)= \varepsilon.\cG_J(t=i B,y^1=1,y^2=0)\ ,
\ee
and $\varepsilon$ is a polarization tensor. The operator $ \overline{\varepsilon.\cG_J}$ is the Rindler reflection of the operator $\cG_J$, defined via 
\be\label{obbar}
\overline{\varepsilon.\cG_J(B)}= \overline{\varepsilon}.\cG_J^\dagger(t=i B,y^1=-1,y^2=0)\ ,
\ee
where the Hermitian conjugate on the right-hand side does not act on the coordinates (see \cite{Hartman:2016lgu} for a detailed discussion on the Rindler reflection), and $\overline{\varepsilon}$ is the Rindler reflection of the polarization $\varepsilon$:
\be
\overline{\varepsilon^{\mu \nu \cdots}} \equiv (-1)^{P}(\varepsilon^{\mu \nu \cdots})^*
\ee
The parameter $P$ is the number of $t$-indices plus $y^1$-indices.  

\subsubsection*{Regge Limit}
Following \cite{Afkhami-Jeddi:2016ntf}, we parametrize the coordinates
\be\label{uv}
u=\frac{1}{\sigma}\ , \qquad v=-\sigma B^2 \rho
\ee
with $B>0, \sigma>0$ and $0< \rho <1$. The Regge limit is defined as
\be\label{reggelimit}
\sigma \rightarrow 0 
\ee
with $\rho, B$ fixed.

\subsubsection*{Statement of Causality}

Let us first write the correlator as $F=1+\delta F$. A CFT is causal if and only if $\delta F$ obeys the following conditions in the Regge limit \cite{Afkhami-Jeddi:2016ntf} :
\begin{itemize}
\item {$\text{Im}(\delta F)$ does not grow faster than $1/\sigma$,}
\item{$\text{Im} (\delta F) \le 0$}
\end{itemize}
which are precisely the chaos growth and sign bounds of \cite{Maldacena:2015waa}. It is important to note that the above constraints are applicable only when $\delta F$ is  perturbatively small. This happens naturally for CFTs with large central charge.\footnote{For generic CFTs, $\delta F$ is perturbatively small in the lightcone limit $\rho\rightarrow 0$. Hence, in the lightcone limit causality also imposes non-trivial constraints \cite{Hartman:2015lfa, Hartman:2016lgu,Hartman:2016dxc,Hofman:2016awc}.} We will be using these bounds and discussing their applicability in the remainder of this paper.

\subsection{A Simple Constraint on Higher Spin Particles}
\label{sec:SimpleConstraintHigherSpin}

Now let us discuss the simplest example of a causality constraint on higher spin particles. In order to do that we consider {\it holographic CFTs} -- CFTs with large central charge and a sparse spectrum, in $d\ge 3$ spacetime dimensions. Furthermore, let us make the strong assumption that $\cG_J$ and $\CO$ are  primary operators dual to a massive higher spin particle and a scalar particle in AdS, and that the only interactions in this AdS theory (at least at this order in perturbation theory, or at energies below the gap) are due to gravity \cite{Afkhami-Jeddi:2016ntf,Afkhami-Jeddi:2017rmx,Afkhami-Jeddi:2018own}.  Then we can write the conformal block decomposition of the correlator from equation (\ref{maing}) as
\begin{align}\label{defFSpectator}
F = \blockS{\cG_J}{\cG_J}{\CO}{\CO}{1} &+
\blockS{\cG_J}{\cG_J}{\CO}{\CO}{T}
 \\
& + \blockT{\cG_J}{\cG_J}{\CO}{\CO}{\sum [\CO \CO]} 
+ \blockT{\cG_J}{\cG_J}{\CO}{\CO}{\sum [\cG_J \cG_J]}  + \cdots \nonumber
\end{align}
where each diagram indicates a set of contributing conformal blocks, and the ellipsis denotes higher order gravitational interactions.  These diagrams are simply the conformal block decomposition of a bulk Witten diagram involving a single graviton exchange between $\cG_J$ and $\CO$.

For simplicity, let us now take $\CO$ to be a heavy operator: $\Delta_\CO\gg \Delta_{\cG_J}$. This allows us to ignore the third set of conformal blocks in (\ref{defFSpectator}). Causality of this simplified correlator was studied in \cite{Afkhami-Jeddi:2018apj} which ruled out all operators $\cG_J$ with $J>2$ (see section 3 of \cite{Afkhami-Jeddi:2018apj}). Let us briefly sketch the argument of  \cite{Afkhami-Jeddi:2018apj}. First, we smear $F$ following \cite{Afkhami-Jeddi:2016ntf} in such a way that it projects out the double trace contributions of $\cG_J$ without spoiling the Rindler reflection symmetry of the correlator (see appendix \ref{smearing})
\bea\label{smeared}
\delta F_{\text{smeared}}=\blockS{\cG_J^{\text{smeared}}}{\cG_J^{\text{smeared}}}{\CO}{\CO}{T} +\cdots\ .
\eea
After smearing, $\delta F_{\text{smeared}}$ is a function of $\rho$, various OPE coefficients, and polarization of the operator $\cG_J$. It was argued in   \cite{Afkhami-Jeddi:2018apj} that that the condition $\text{Im} (\delta F_{\text{smeared}}) \le 0$ for all polarizations of the operator $\cG_J$ cannot be satisfied in the limit $\rho\rightarrow 1$ for $J>2$.

\section{Causality Constraints for Composites in AdS?}
\label{sec:ConstraintsinAdS}

In section \ref{sec:SimpleConstraintHigherSpin} we summarized a result from \cite{Afkhami-Jeddi:2018apj}, which appears to put very strong constraints on the existence of elementary higher spin particles in AdS/CFT.  But higher spin \emph{composite} particles are ubiquitous in physics -- indeed we ourselves are higher spin `particles'!  So from the point of view of these constraints, we would like to investigate in what sense composite and fundamental particles are different.

We will begin with the simplest possible case, and understand why the constraint does not apply to two-particle states in AdS, which can be viewed as `bound states' due to the AdS curvature.  Two-particle states are represented in the CFT as double-trace operators with arbitrarily large spin, so we will study correlators involving a pair of these double-trace operators.   We will see that the causality bound doesn't apply because there are extra contributions to these correlators as compared to the case of fundamental higher spin particles.  This result should also apply to other weakly bound states, explaining why the existence of hydrogen atoms is not constrained by causality.

It would be surprising if causality bounds apply to unstable particles with short lifetimes.  We will  discuss this issue in the context of AdS/CFT, where unstable bulk particles are  dual to CFT operators that  include large admixtures of multi-trace operators.  These effects  may complicate or eliminate the causality bounds.

\subsection{No Constraint on Free Two-Particle States in AdS}
\label{sec:NoConstraintFreeParticles}

In this section we will demonstrate that free two-particle states bound only by the effect of the AdS curvature are not constrained by causality.  Physically, this result seems very obvious, but the goal is to establish it from the point of view of correlators and conformal blocks.

We study  a simple toy model  in AdS, working in $(4+1)$-dimensions as it will simplify some algebraic expressions. We consider two free scalar fields $\phi_1$ and $\phi_2$ in AdS which are dual to two primary scalar operators $\O_1$ and $\O_2$ with dimensions $\Delta$. There are various double trace operators in this theory such as $[\O_1 \O_1]_{n,\ell}$, $[\O_2 \O_2]_{n,\ell}$, $[\O_1 \O_2]_{n,\ell}$. We are interested in the mixed double trace operator of dimension $2\Delta+2n+\ell$ and spin $\ell$ which can be schematically written 
\be
[\O_1 \O_2]_{n,\ell}\sim   \O_1\Box^n \partial_{\mu_1}\partial_{\mu_2}\cdots \partial_{\mu_\ell} \O_2+\cdots
\ee
with the exact expressions in appendix \ref{app_dt}. Our goal is to show that the argument of section \ref{sec:SimpleConstraintHigherSpin} when applied to $[\O_1 \O_2]_{n,\ell}$ does not lead to a constraint for any $\ell$.

\subsubsection{Four-Point Function}
We start with the correlator $\langle [\O_1 \O_2]_{0,\ell} [\O_1 \O_2]_{0,\ell} \psi \psi\rangle$ for $\ell=3$, where $\psi$ is a heavy scalar that only interacts with $\phi_1$ and $\phi_2$ through gravity (this bulk theory is manifestly dual to a CFT with large central charge and a sparse spectrum).  This correlator can be straightforwardly computed from the bulk Witten diagrams of figure \ref{dt1}, and we find
\begin{align}\label{stark}
\frac{\langle \psi(x_1) \varepsilon_2\cdot \O_3(x_2)\varepsilon_3\cdot\O_3(x_3) \psi(x_4)\rangle}{\langle \psi(x_1)   \psi(x_4)\rangle}= \langle  \varepsilon_2\cdot\O_3(x_2)\varepsilon_3\cdot\O_3(x_3) \rangle + 
\frac{\mathcal{D} [\tilde{G}]}{\langle \psi(x_1)   \psi(x_4)\rangle}\ ,
\end{align}
where $\O_3 \equiv [\O_1 \O_2]_{0,\ell=3}$ (see equation (\ref{spin3})) and $\varepsilon\cdot \O_3\equiv \varepsilon_{\mu_1}\varepsilon_{\mu_2}\varepsilon_{\mu_3}\O_3^{\mu_1\mu_2\mu_3}$. Correlator $\tilde{G}$ in the above equation is the partially connected six-point function
\begin{align}
\tilde{G}=\langle \psi(x_1) \O_1(x_2)\O_1(x_3) \psi(x_4)\rangle_h &\langle\O_2(x_2')\O_2(x_3') \rangle\nonumber\\
&+\langle \psi(x_1) \O_2(x_2')\O_2(x_3') \psi(x_4)\rangle_h \langle\O_1(x_2)\O_1(x_3) \rangle
\end{align}
where, subscript $h$ stands for the graviton exchange Witten diagram and the operator $\mathcal{D}$ can be obtained from equation (\ref{spin3})
\ba
\mathcal{D}=N_3^2  \lim_{x_2'\rightarrow x_2,  x_3'\rightarrow x_3}\left(D_1^3 -\frac{6}{\Delta} D_0^2 D_1\right)  \left({D'}_1^3 -\frac{6}{\Delta} {D'}_0^2 {D'}_1\right)
\ea
with 
\ba
D_1= (\varepsilon_2.\partial_2-\varepsilon_2.\partial_{2'})\ , \qquad D_0^2= \varepsilon_2.\partial_2 \varepsilon_2.\partial_{2'}\ , \nonumber\\
D'_1= (\varepsilon_3.\partial_3-\varepsilon_3.\partial_{3'})\ , \qquad {D'}_0^2= \varepsilon_3.\partial_3 \varepsilon_3.\partial_{3'}\ .
\ea

\subsubsection{Smeared Regge Correlator and Causality}

To simplify calculations, let us further restrict to $\Delta=2$ in $d=4$. Since $\psi$ is a heavy scalar operator, we can use the Regge OPE of $\psi\psi$ \cite{Afkhami-Jeddi:2017rmx} to explicitly compute the Regge correlator (see appendix \ref{app_hhll})
\be\label{scalar4}
\frac{\langle \psi(x_1) \O_1(x_2)\O_1(x_3) \psi(x_4)\rangle_h}{\langle \psi(x_1)   \psi(x_4)\rangle\langle\O_1(x_2)\O_1(x_3) \rangle}=\frac{\langle \psi(x_1) \O_2(x_2)\O_2(x_3) \psi(x_4)\rangle_h}{\langle \psi(x_1)   \psi(x_4)\rangle\langle\O_2(x_2)\O_2(x_3) \rangle}=-i \frac{80 \Delta_\psi}{c_T \pi^3} \frac{z \bz}{(z+\bz)^3}\ ,
\ee
where, $c_T$ is the central charge of the dual CFT. Now, we choose the points (\ref{points}) and take the Regge limit (\ref{reggelimit}).

Then we choose the following (null) polarizations for $\O_3$
\be 
\varepsilon_2=(1,\xi ,i \lambda ,\lambda)\ , \qquad \varepsilon_3=(-1,-\xi ,-i \lambda ,\lambda)\ ,
\ee
where, $\xi=\pm 1$. It is easy to compute the smeared two-point  function following appendix \ref{smearing}
\be
\int d\tau d\vec{y}\langle \O_3(x_2) \O_3(x_3)\rangle=\frac{\pi ^2 \left(56 \lambda ^6+105 \lambda ^4+60 \lambda ^2+10\right)}{524288}\ .
\ee

Next we use the scalar correlator (\ref{scalar4}) to compute the Regge correlator (\ref{stark}) of $\O_3$. After performing the smearing for $d=4$ and $\Delta=2$, in the limit $\rho\rightarrow 1$, we finally obtain
\be\label{smeared_dt}
\frac{\int d\tau d\vec{y} \langle \psi(x_1) \varepsilon_2\O_3(x_2)\varepsilon_3.\O_3(x_3) \psi(x_4)\rangle}{\langle \psi(x_1)   \psi(x_4)\rangle\int d\tau d\vec{y}\langle \O_3(x_2) \O_3(x_3)\rangle}\approx 1-\frac{10 i \Delta_\psi\left(98 \lambda ^6+343 \lambda ^4+405 \lambda ^2+190\right)}{7 \pi ^3 \sigma  c_T\left(56 \lambda ^6+105 \lambda ^4+60 \lambda ^2+10\right) }\ .
\ee
Note that Im$(\delta F)$ is negative which is already consistent with causality.

\subsubsection{Comparison Between Single Trace and Double Trace Operators}

Let us now point out certain key differences between  the double trace result (\ref{smeared_dt}) and the single trace results of \cite{Afkhami-Jeddi:2018apj}.
\begin{itemize}
\item{First of all, note that the smeared correlator $\delta F_{\text{smeared}}$ is finite in the limit $\rho\rightarrow 1$. This limit corresponds to a high energy scattering deep into the bulk and in this limit, smeared correlators of single trace operators have singularities. In particular,  $\delta F_{\text{smeared}}$ for a single trace primary operator of spin $\ell$ in the limit $\rho\rightarrow 1$ has the following form   \cite{Afkhami-Jeddi:2018apj}
\be\label{singular}
\delta F_{\text{smeared}} \sim -i \frac{P(\lambda)}{c_T  \sigma (1-\rho)^{d+2\ell-3}}\ 
\ee
which grows even for $\ell=0$. Whereas, the smeared correlator of double trace operators (\ref{smeared_dt}) is finite in the limit $\rho\rightarrow 1$ for spin $\ell=3$. It is not difficult to see that the same is true for all free two-particle states (with or without spin) bound only by the effect of the AdS curvature. First, note that for the single trace operator $\O_1$ or $\O_2$, the singularity at $\rho=1$ in (\ref{singular}) comes from the following volume integral (for any $\Delta$ in $d=4$), which is approximately
\be
\delta F_{\text{smeared}} \sim \int \frac{d\tau d^2\vec{y}}{\tau^2+\vec{y}^2}
\ee 
at large $\tau^2+\vec{y}^2$. Whereas, for the double trace operator $[\O_1\O_2]_{0,0}$ the above volume integral  at $\rho=1$ and at large $\tau, \vec y$ becomes
\be
\delta F_{\text{smeared}} \sim \int \frac{d\tau d^2\vec{y}}{(\tau^2+\vec{y}^2)^{1+\Delta}}
\ee 
which is finite when $\Delta$ satisfies the unitarity bound. For double trace operators with $n,\ell>0$, the above integrand schematically has the following structure
\be
 \partial \partial \cdots \partial \ \frac{1}{(\tau^2+\vec{y}^2)^{1+\Delta}} \nonumber
\ee 
which decays at least as fast as the scalar case. Disappearance of the singularity at $\rho=1$ for free two-particle states signifies that the two-particle states do not have  hard centers and hence it is expected that the same feature persists in any $d\ge 3$. 
}
\item{$P(\lambda)$ in the above expression is a polynomial in $\lambda^2$ and causality requires that each power of $\lambda^2$ in  $P(\lambda)$ must be individually positive \cite{Afkhami-Jeddi:2016ntf}. For single trace operators, coefficients of individual powers of $\lambda^2$ generally change sign -- this leads to non-trivial constraints on the OPE coefficients. However, for double trace operators, each power of $\lambda^2$ in  $P(\lambda)$ has the same sign which implies that causality does not impose any non-trivial constraints.}
\item As we discuss in detail in section \ref{sec:HydrogenAllowed} below, the absence of constraints can be traced to the presence of certain double-trace operators made from the constituents of the two-particle state we have been studying.
\end{itemize}

\subsection{Why Hydrogen Atoms in AdS Aren't Ruled Out}\label{hydrogen}
\label{sec:HydrogenAllowed}

In the last section we showed that causality constraints do not apply to generalized free theory double trace operators like 
\be
[\O_1 \O_2]_{n,\ell}\sim   \O_1\Box^n \partial_{\mu_1}\partial_{\mu_2}\cdots \partial_{\mu_\ell} \O_2+\cdots\ .
\ee
Now let's imagine adding interactions that bind particle $1$ and $2$ together in AdS, forming analogs of `hydrogen atoms'.  Such interactions will give these states anomalous dimensions (corresponding to binding energies), and mix them with states involving the dual of the bulk force carrier.  In the CFT language, it is natural to ask:  to what extent can we approximate the primary double-trace operator $[\O_1 \O_2]$ as a single-trace\footnote{The conventional $n$-trace terminology is a bit artificial here; really we are asking to what extent composite particles in AdS behave as though they are fundamental.} primary?  

This question can be addressed by considering a four-point function $\langle \psi \psi [\O_1 \O_2][\O_1 \O_2]\rangle$ in a CFT with large central charge $c_T$  and a sparse spectrum of higher spin operators. At the leading order in $1/c_T$, the t-channel expansion of this four-point function necessarily receives  contributions from the following conformal blocks:
\begin{align}\label{dtrace}
\langle \psi \psi [\O_1 \O_2] [\O_1 \O_2 ]\rangle&=\blockS{[\O_1 \O_2]}{ [\O_1 \O_2] }{\psi}{\psi}{1}
+
\blockS{[\O_1 \O_2] }{[\O_1 \O_2] }{\psi}{\psi}{T}
\nonumber\\
&+\blockTT{[\O_1 \O_2] }{[\O_1 \O_2] }{\psi}{\psi}{\sum[\CO_1 \CO_1]}+\blockTT{[\O_1 \O_2] }{[\O_1 \O_2] }{\psi}{\psi}{ { \sum [\CO_2 \CO_2]}}\nonumber\\
&+\blocklarge{[\O_1 \O_2] }{[\O_1 \O_2] }{\psi}{\psi}{ { \sum [[\O_1 \O_2] [\O_1 \O_2] ]}}+\cdots\ ,
\end{align} 
where  we have ignored contributions from $[\psi\psi]$ because $\psi$ is heavy.  We have also assumed that all cubic interactions between $\phi_1$ and $\phi_2$ are small (since such couplings are not obligatory). The first line in equation (\ref{dtrace}) always contributes to the four-point function $\langle \psi \psi [\O_1 \O_2] [\O_1 \O_2] \rangle$. 

On the other hand, the relative strength between the second and the third line depends on the type of interactions between bulk fields $\phi_1$ and $\phi_2$.   Moreover, the  presence of the second line is particular to correlators of composite states.  The terms in the second line have arbitrarily large spin, and thus they can easily compete with the stress-tensor exchange block on the first line of equation (\ref{dtrace}).

\begin{figure}[h!]
\centering
\includegraphics[scale=0.36]{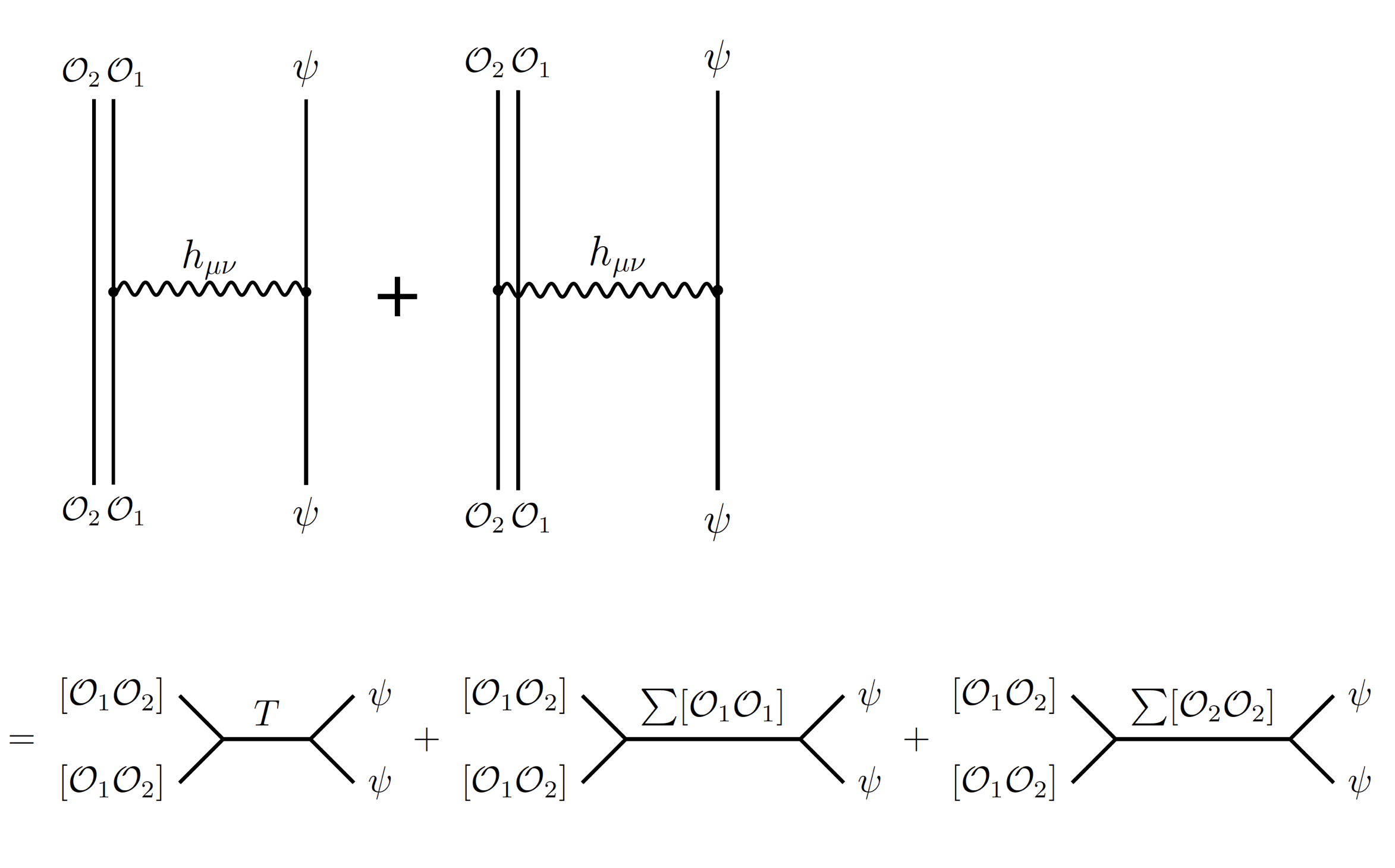}
\caption{ \label{dt1} \small The correlator $\langle \psi \psi [\O_1 \O_2] [\O_1 \O_2] \rangle$ when bulk fields $\phi_1$ and $\phi_2$ are free or weakly interacting is completely determined from the above Witten diagrams. For $[\O_1 \O_2]_{n,\ell}$ with nonzero $n$ and/or $\ell$, the left hand side should be acted on by the appropriate derivative operator. }
\end{figure}
Let us consider two extreme scenarios. First, if the bulk fields $\phi_1$ and $\phi_2$ are free or weakly interacting, the four-point function can be approximated by the Witten diagrams \ref{dt1} and hence we can neglect the third line of equation (\ref{dtrace}). In this case, as we saw in section \ref{sec:NoConstraintFreeParticles}, there are no causality constraints. In CFT language, the causality violations are avoided because of the exchanges of $[\CO_1 \CO_1]$ and $[\CO_2 \CO_2]$  in the conformal block decomposition of equation (\ref{dtrace}).

\begin{figure}[h!]
\centering
\includegraphics[scale=0.36]{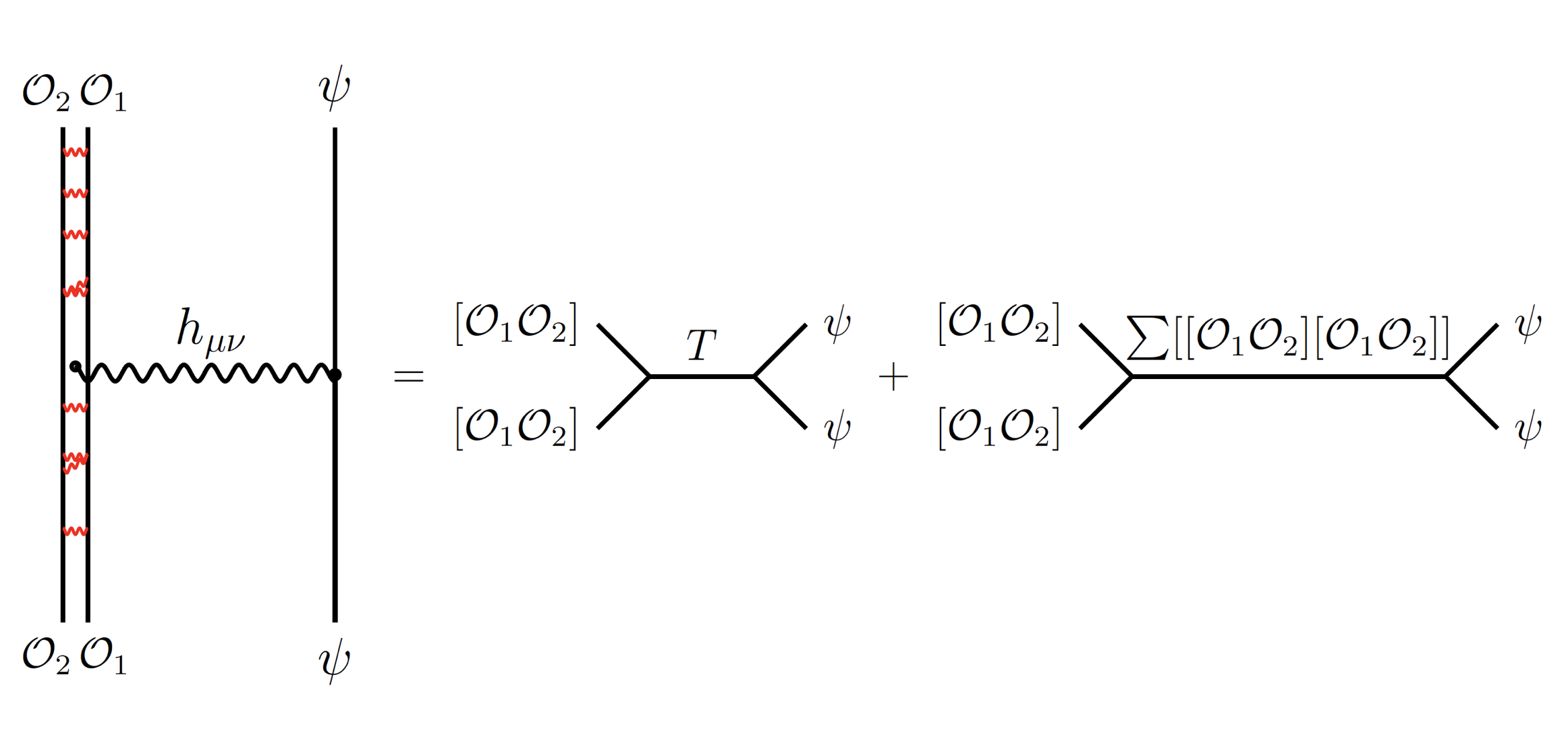}
\caption{\small The double trace operator $[\O_1 \O_2] $ in the correlator $\langle \psi \psi [\O_1 \O_2] [\O_1 \O_2] \rangle$ effectively behaves like a single trace primary when bulk fields $\phi_1$ and $\phi_2$ are strongly interacting.}\label{dt2}
\end{figure}
On the other hand, if bulk fields $\phi_1$ and $\phi_2$ are strongly interacting then we can no longer treat $\O_1$ and $\O_2$ individually, as shown in figure \ref{dt2}. In this case the last line of equation (\ref{dtrace}) dominates over the second line, implying $[\O_1 \O_2] $ can be approximated as a single trace primary operator. Glueballs and mesons of confining large $N$ gauge theories belong to this class and hence the argument of \cite{Afkhami-Jeddi:2018apj} still applies. Whereas, hydrogen atoms obviously are more similar to the scenario of figure \ref{dt1} and hence hydrogen atoms with spin more than two are not ruled out in AdS. 

One might still wonder what happens if, beginning with a Hydrogen-like bound state, we increase the strength of the coupling holding the constituents together.  Is there a transition to a regime where the causality bounds apply?  
In fact, we expect that even for order-one couplings, Hydrogen-like composites  will not be constrained by the causality bound, as the operators from the second line of equation (\ref{dtrace}) can still contribute.  Thus we only expect  causality constraints to apply to bound states that are parametrically lighter than their constituents.  Confining theories are an extreme example, since quarks and gluons do not exist as finite-energy states at all.  But our bounds may help to rule out proposals for parametrically light bound states with large spin.  In this sense, our bounds may be viewed as an extension of the Weinberg-Witten theorem \cite{Weinberg:1980kq}.

\subsection{Unstable Particles and Operator Mixing}
\label{sec:OperatorMixingandUnstableParticles}

On physical grounds, we might not expect causality constraints to be  applicable to `particles' with very short lifetimes, because these particles may decay before any causality violation can be unambiguously detected.  Let us see how particle instability in AdS manifests in the dual CFT, and how it might affect causality constraints.

When we study a completely free QFT in AdS, the single particle ground states correspond to CFT primaries (which are typically denoted `single-trace' operators).  Multi-particle states with fixed particle number can also be organized into primaries and descendants as well.  Once AdS interactions are turned on, states with different numbers of particles  mix, and at finite coupling the exact CFT primaries will not have definite particle number.  This effect becomes especially pronounced in the presence of unstable particles, which devolve into admixtures\footnote{Similar observations would apply  whenever a QFT lives in a compact space and thus has a discrete spectrum.  The only feature unique to AdS is the decomposition of states in terms of primaries and descendants, which is a consequence of conformal symmetry.} dominated by their decay products.

In order to illustrate these effects,  let us  imagine a toy bulk theory with two scalar fields and an action
\be
S = \int d^{d+1} X \left( \frac{1}{2} (\nabla \Phi)^2  + \frac{1}{2} (\nabla \chi)^2 - \frac{1}{2} M^2 \Phi^2 - \frac{1}{2} m^2 \chi^2 - \frac{g}{2} \Phi \chi^2  \right)
\ee
where we can vary $M$ and $m$, allowing us to study the case  $\Delta_\Phi > 2 \Delta_\chi$ where $\Phi \to 2 \chi$ decays are allowed.  When $g=0$ the boundary dual is a theory of generalized free fields $\CO_\Phi$ and $\CO_\chi$, whose spectra include double and multi-trace operators.

Our  interest is in the decomposition of primaries at $g \neq 0$ into the Hilbert space of $g=0$ states.  In particular, we would like to understand to what extent $\CO_\Phi$ mixes with double-trace operators $[\CO_\chi \CO_\chi]_{n, \ell}$.  Since $\CO_\Phi$ is a scalar, and perturbation theory preserves angular momentum, $\CO_\Phi$ can only mix with the $\ell = 0$ double traces. Via the operator/state correspondence, we can study mixing of operators by considering the perturbative mixing of states, and vice-versa.  Eigenstates of the global AdS  Hamiltonian  correspond to CFT states with definite scaling dimension.

To study operator mixing effects, it's  convenient to use old-fashioned perturbation theory, as it provides a formula for the mixing of Hamiltonian eigenstates.
Adapting to the CFT context \cite{Katz}, the textbook formula becomes
\be
\CO_{\Phi} = \CO_{\Phi_0} + \sum_n \frac{\langle \CO_{\Phi_0} |  V | [\CO_\chi \CO_\chi]_{n, 0} \rangle}{\Delta_\Phi - (2 \Delta_\chi + 2n) }   [\CO_\chi \CO_\chi]_{n, 0} + \cdots
\ee
where the operators on the right hand side create states in the $g=0$ Hilbert space.  The $\CO_\Phi$ on the left-hand side is the exact (or at least perturbative in $g$) primary.

When the denominator is order $1$, as is possible when the $\Phi \to 2 \chi$ decay channel is open, we may have a very large\footnote{The singularity is unphysical; it comes from expanding a formula like $\sqrt{(M^2-4m^2)^2 + g^4}$ in small $g$. Near the singularity we must instead use degenerate perturbation theory.} mixing.  This is because when we interpret CFT dimensions as bulk energies,  operator dimensions naturally have units of $1/R_{\text{AdS}}$.  If for example we are in AdS$_4$, then $g$ has units of energy, and so we could be in the regime $M \gtrsim m \gg g  \gg 1/R_{\text{AdS}}$.  In this case we would still say that we have a very weakly coupled bulk field theory, since the coupling is small compared to the masses of the particles.  But the mixing effect for unstable particles would be very large, so that the true primary $\CO_\Phi$ behaves very differently from the naive generalized free field $\CO_{\Phi_0}$.  In this regime  correlators of $\CO_\Phi$ may behave almost exactly like correlators of $[\CO_\chi \CO_\chi]_{n, 0}$, which are automatically free from causality constraints.

\subsubsection*{Quantitative Estimate}

Now let us estimate to what extent operator mixing alleviates causality bounds.  Our goal will be to determine how small the mixing must be to be confident that causality bounds apply.  

We will parameterize the exact primary as
\be
\CO = \frac{1}{\sqrt{1 + \alpha^2}} \CO_0 +  \frac{\alpha}{\sqrt{1 + \alpha^2}} [\CO \CO]
\ee
where $[\CO \CO]$ denotes a combination of double-trace operators, and we have chosen this representation to keep $\CO$ normalized.  When the mixing $\alpha \gg 1$, correlators of $\CO$ will be dominated by the double-traces, which manifestly preserve causality.

When we study the correlator $F = \langle \CO  \psi \psi \CO\rangle$ and smear as described in section \ref{sec:SimpleConstraintHigherSpin}, we obtain  two contributions which take the parametric form
\be
\delta F_{\rm smeared} \sim \frac{i \Delta_\psi \Delta_\CO}{c_T \sigma (1 + \alpha^2)} \left(\alpha^2+\sum_{i=0}^p \frac{1}{(1 - \rho)^{d - 3 +i}} +\text{finite part}   \right)
\ee
where $p=0$ for scalar external operators, however, if operator $\CO$ has spin $\ell$ then $p=2\ell$.  In the case $d=3$ and $p=0$ the power-law is replaced with a $\log(1 - \rho)$.  For the theory to remain under control, we must take $1 - \rho > \frac{1}{\Delta_{\text{gap}}}$.  This means that mixing effects do not influence causality constraints only if
\be
\alpha^2 \ll \log(\Delta_{\rm gap})
\ee
in the case $d=3$.  Typically $\alpha$ will grow with $R_{\text{AdS}}$, so these mixing effects obstruct the flat space limit of the AdS causality bounds.  Physically, this result has a very simple interpretation:  we cannot obtain causality bounds from particles that  decay before they can scatter.

\section{Constraining a Gauge Theory Inside the AdS Bulk }
\label{sec:AdSAnalysis}

In this section we will place a confining large $N$ gauge theory \emph{inside AdS} and study the implications for causality in the dual boundary CFT.  Note that when we refer to `the gauge theory' or `$N$' we will always be referring to a $d+1$ dimensional gauge theory in AdS$_{d+1}$, and not to the boundary CFT$_d$.  We focus on $d=3$, as we do not expect confinement in higher dimensions, and causality constraints on higher spin particles do not apply in lower dimensions. The central charge of the CFT$_3$ will be related to the AdS scale and bulk Planck  scale by \cite{Myers:2010tj}
\be\label{ct}
c_T=\frac{24 R_{\text{AdS}}^2 M_{Pl}^2}{\pi^2}\ 
\ee
and is a priori completely unrelated to the $N$ of the bulk gauge theory.  Our AdS/CFT setup only includes the gauge theory, gravity, and perhaps some spectator fields in the bulk of AdS below a cutoff scale $\Lambda_{\text{UV}}$.

A bulk gauge theory might be in either a confining or Coulomb phase, and the difference will have a marked effect on the spectrum of  the boundary CFT.  In the Coulomb phase, free charges have finite energy, and so the bulk gauge field will act on the vacuum to create finite-energy gluons.  Such a bulk gauge field would then be dual to a conserved current in the boundary CFT.   However, in the confining phase bulk colored states will have infinite energy, decoupling from the spectrum.  In particular, if the bulk gauge theory confines, then the boundary CFT will not include any symmetry currents\footnote{Since AdS acts as an IR regulator \cite{Callan:1989em}, we expect that the confinement scale $\Lambda_{\text{QCD}}$ cannot be much smaller than the AdS scale $\frac{1}{R_{\text{AdS}}}$ without transitioning back to the Coulomb phase.  Conversely, bulk gauge fields and their dual currents in the CFT cannot be too strongly coupled in the Coulomb phase.  We can interpret bootstrap bounds \cite{Dymarsky:2017xzb} on the maximum current OPE coefficients $\langle JJJ \rangle$ as indicative of this transition. } dual to the bulk gauge field. 

We will be studying both the large $N$ limit of our bulk gauge theory and the large $c_T$ limit of the boundary CFT.   In more physical terms, we will imagine fixing $R_{\text{AdS}}$ and $\Lambda_{\text{QCD}}$ while taking $N, \frac{M_{Pl}}{\Lambda_{\text{QCD}}} \to \infty$.  This means that  glueballs and mesons will have lifetimes much larger than the $R_{\text{AdS}}$ timescale.  So we will not need to worry about meson or glueball decays, or mixing of single and multi-trace operators.  Note that this scaling does not commute with the flat space limit, as in flat space, for any finite but large $N$, hadrons would have a finite lifetime.  We will be interested in comparing the gravitational and $1/N$ couplings, and so we will choose scalings so that these couplings are similarly tiny.

In this section, we derive a bound on $N$ under the assumption of confinement of the bulk gauge theory and the presence of higher spin glueballs/mesons. The CFT based argument has several advantages. First of all, as summarized in section \ref{sec:BriefReviews}, the statement of causality is well understood in CFT \cite{Hartman:2015lfa,Hartman:2016dxc,Hofman:2016awc,Hartman:2016lgu,Afkhami-Jeddi:2016ntf}. It provides a condition on how certain four-point functions in CFT must behave. It was shown in \cite{Afkhami-Jeddi:2018apj} that single trace primary operators with spin $J>2$ violate this causality constraint, ruling out elementary higher spin particles in AdS. 

The argument of \cite{Afkhami-Jeddi:2018apj} is not obviously applicable for CFT operators $\cal{G}$ and $\Pi$ (with spin $J>2$) which are dual to the glueball $G$ and meson $\pi$, respectively.  In our setup, the traditional notion of `single-trace' vs `double-trace' primary operator is not so well-defined.  One might take the viewpoint that the operators $\cal{G}$ and $\Pi$ are double-trace, as a consequence of the fact that mesons and glueballs are not elementary particles. However, in the strict limit of $N\rightarrow \infty$, we will argue that operators $\cal{G}$ and $\Pi$ behave exactly like `single-trace' operators and hence the bound of \cite{Afkhami-Jeddi:2018apj} should be applicable. Moreover, a simple estimation of $1/N$ contributions for large but finite $N$ implies that causality in the dual CFT can only be restored if the bound (\ref{bound}) is satisfied.

\subsection{Bound from the Dual CFT}

We consider CFT operators $\cal{G}$ and $\Pi$ which are dual to glueballs $G$ and mesons $\pi$, respectively.  We will assume that gauge theories contain infinite towers of $G$'s and $\pi$'s with all spins and hence the dual operators $\cal{G}$ and $\Pi$ must also come in infinite towers.  For simplicity, we will also include a colorless bulk scalar field $\phi$ dual to a scalar CFT primary $\cal O$, which we assume only interacts with the gauge sector via gravity.  We will obtain bounds with and without $\phi$, but its inclusion provides a simple and  stark demonstration of how causality bounds constrain gauge theories coupled to gravity.

In AdS, we also have a bulk graviton $h$ that couples to $G$, $\pi$, and $\phi$ with parametric strength $\sqrt{G_N}$ fixed by the equivalence principle. In the CFT side this implies that the three-point functions with the stress tensor $T$ are suppressed by the central charge 
\be
\frac{\langle {\cal G} {\cal G} T\rangle}{\sqrt{\langle T T \rangle}} \sim \frac{1}{\sqrt{c_T}}\ , \qquad \frac{\langle \Pi \Pi T\rangle}{\sqrt{\langle T T \rangle}} \sim \frac{1}{\sqrt{c_T}} \ , \qquad \frac{\langle {\cal O} {\cal O}  T \rangle}{\sqrt{\langle T T \rangle}}  \sim \frac{1}{\sqrt{c_T}}
\ee
where $c_T$ is the central charge defined from the stress tensor two-point function \cite{Osborn:1993cr,Erdmenger:1996yc}.\footnote{Note that this $T_{\mu \nu}$ is the CFT stress tensor and hence it is different from the $T_{\mu\nu}$ appeared in equation (\ref{eq:GGTScaling}). In $(2+1)$ dimensions, the two-point function of the CFT stress tensor $T$ is given by
\be
\langle T_{\mu \nu}(x)T_{\rho \sigma}(0)\rangle =\frac{c_T}{2x^{6}}\left(I_{\mu \rho}(x)I_{\nu \sigma}(x)+I_{\mu \sigma}(x)I_{\nu \rho}(x) -\frac{2}{3}\eta_{\mu \nu}\eta_{\rho \sigma}\right)\ ,
\ee
where $I_{\mu \nu}(x)$ is completely fixed by conformal invariance 
\be
I_{\mu \nu}(x)=\eta_{\mu \nu}-\frac{x_\mu x_\nu}{x^2}\ .
\ee
}
We are at the holographic limit $c_T\gg 1$ with a sparse spectrum.  From the perspective of the dual CFT, there are two small parameters: $1/N$ and $1/c_T$ -- a priori these are independent parameters. However, next we will argue that the scenario $N\gg \sqrt{c_T}$ leads to violations of causality.

\subsection{Constraints on Large $N$ Gauge Theories in AdS}

Now let us discuss how causality constrains large $N$ gauge theories in AdS.  We begin with a warm-up example involving a spectator field interacting with a high-spin glueball or meson, and then discuss interactions between scalar and high-spin hadrons.

\subsubsection*{High-Spin Glueball and a Spectating Scalar}

Consider a simple Regge correlator involving two higher spin glueball operators $\cG_J$ and two spectating scalar operators $\CO$.  Since the gauge sector can only interact with $\CO$ via gravity, this correlator can be approximated as \cite{Afkhami-Jeddi:2016ntf,Afkhami-Jeddi:2017rmx,Afkhami-Jeddi:2018own}
\begin{align} \label{defFSpectator1}
F &= \blockS{\cG_J}{\cG_J}{\CO}{\CO}{1} +
\blockS{\cG_J}{\cG_J}{\CO}{\CO}{T}
\nonumber \\
& + \blockT{\cG_J}{\cG_J}{\CO}{\CO}{\sum [\CO \CO]} 
+ \blockT{\cG_J}{\cG_J}{\CO}{\CO}{\sum [\cG_J \cG_J]}  + \cdots
\end{align}
where each diagram indicates a set of contributing conformal blocks, and the ellipsis denotes higher order gravitational interactions.  These diagrams are simply the conformal block decomposition of a bulk Witten diagram involving graviton exchange between $\cG_J$ and $\CO$.

We can further simplify by making $\CO$ heavy: $\Delta_\CO\gg \Delta_{\cG_J}$. This allows us to ignore the third set of conformal blocks in (\ref{defFSpectator1}). 
The correlator (\ref{defFSpectator1}) is now identical to the correlator (\ref{defFSpectator}) implying that the bound of  \cite{Afkhami-Jeddi:2018apj} is applicable here as well. We again smear $F$ following \cite{Afkhami-Jeddi:2016ntf} in order to project out the double trace contributions of $\cG_J$ 
\bea 
\delta F_{\text{smeared}}=\blockS{\cG_J^{\text{smeared}}}{\cG_J^{\text{smeared}}}{\CO}{\CO}{T} +\cdots\ .
\eea
This correlator, as shown in   \cite{Afkhami-Jeddi:2018apj}, violates causality for any $J>2$.  If we replaced $\cG_J$ with a meson $\Pi_J$ we would obtain the same result.

As discussed in section \ref{sec:OperatorMixingandUnstableParticles}, at large but finite $N$ these bounds may be alleviated by the effects of operator mixing.  If $G_J$ or $\pi_J$ can decay to multi-hadron states, then the corresponding CFT operators $\cG_J$ and $\Pi_J$ will contain large admixtures of multi-trace operators.  In this case the correlator $F$ will receive additional and potentially very important contributions from other multi-trace operators, and the causality bounds may not apply.  Mixing of glueballs will be suppressed by $\frac{1}{N}$, but it may be enhanced by a power of $(\Lambda_{\text{QCD}} R_{\text{AdS}}) \sim \Delta_J$.  This suggests that the bounds may not apply when $N \lesssim \Delta_J$, though the specific dependence will be theory-dependent.

\subsubsection*{Constraints from Bulk Gauge Theory Correlators}

Now let us consider the correlator or AdS scattering amplitude between $\cG_J$ and a spin-zero glueball or meson $\cG_0$ in the Regge limit.  This case is more complicated because these particles' interactions are also mediated by the gauge-theory.

In general, it is not possible to precisely compute  gauge theory correlators in the Regge limit, unless we know more about the gauge theory or its boundary dual.  But we can approximate the correlator using the conformal block decomposition \cite{Afkhami-Jeddi:2016ntf,Afkhami-Jeddi:2017rmx,Afkhami-Jeddi:2018own}
\begin{align}\label{defF}
F&=\blockS{\cG_J}{\cG_J}{\cG_0}{\cG_0}{1}
 \nonumber\\
&+
\blockS{\cG_J}{\cG_J}{\cG_0}{\cG_0}{T}
+\blockT{\cG_J}{\cG_J}{\cG_0}{\cG_0}{ { \sum [\cG_0 \cG_0]}}+\blockT{\cG_J}{\cG_J}{\cG_0}{\cG_0}{ { \sum [\cG_J \cG_J]}}\nonumber\\
&+\blockS{\cG_J}{\cG_J}{\cG_0}{\cG_0}{\sum \cG}+\blockS{\cG_J}{\cG_J}{\cG_0}{\cG_0}{\sum \Pi}+\O\left(\frac{1}{c_T^2}, \frac{1}{N^3} \right)
\end{align}
where, the first term is $\sim N^0 c_T^0$ and the second line is $\sim 1/c_T$. Whereas, the third line of conformal blocks are suppressed by $1/N^2$ and hence they can be ignored in the limit $N\gg \sqrt{c_T}\gg \Delta_{\text{gap}}$.\footnote{Note that in this limit, we can also ignore the mixing effect from section \ref{sec:OperatorMixingandUnstableParticles}.} We  can also choose $\cG_0$ to be  heavy, 
which allows us to ignore\footnote{This isn't strictly necessary, as we could project out  these operators using an additional smearing.  However, since this would add  technical complication,  for simplicity we assume $\cG_0$ are heavy.  } the exchange of $[\cG_0 \cG_0]$ as well. So, we only need to consider the  conformal blocks
\bea\label{delF}
\delta F=\blockS{\cG_J}{\cG_J}{\cG_0}{\cG_0}{T}
 +\blockT{\cG_J}{\cG_J}{\cG_0}{\cG_0}{ { \sum [\cG_J \cG_J]}}+\cdots\ .
\eea
In the bulk, this corresponds to the  graviton exchanged Witten diagram
\be\label{ucontour}
\begin{gathered}\includegraphics[width=0.45\textwidth]{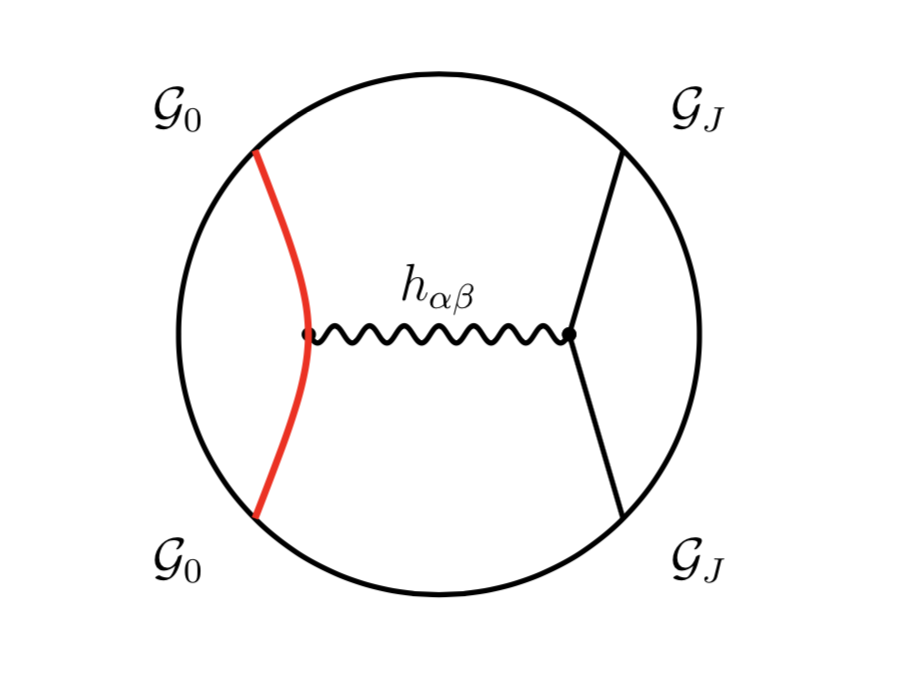}\end{gathered}
\ee
where, on the left side the integral is only over the geodesic that connects the two $\cG_0$ operators, because we assumed that they are heavy. 

\subsubsection*{Bounds}

The operators $\cG_0$ and $\cG_J$  may not be single-trace, however, the correlator (\ref{delF}) is  identical to the correlator of single trace operators considered in \cite{Afkhami-Jeddi:2018apj}.  We smear operators $\cG_J$'s (as shown in appendix \ref{smearing}) to project out the double trace contributions of $\cG_J$ without spoiling the Rindler reflection symmetry of the correlator
\bea\label{smeared12}
\delta F_{\text{smeared}}=\blockS{\cG_J^{\text{smeared}}}{\cG_J^{\text{smeared}}}{\cG_0}{\cG_0}{T} +\cdots\ .
\eea
The resulting smeared correlator is a function of $\rho$, various OPE coefficients, and the polarization of the operator $\cG_J$. This smeared correlator is causal only if it satisfies the condition $\text{Im} (\delta F_{\text{smeared}}) \le 0$ for any polarization of $\cG_J$ and $0\le\rho\le 1$ (see section \ref{sec:SimpleConstraintHigherSpin}). However, it was shown in  \cite{Afkhami-Jeddi:2018apj} that the correlator (\ref{smeared12}) for $J>2$ cannot satisfy the causality condition for all polarizations of $\cG_J$ in the limit $\rho\rightarrow 1$.  However, we  expect that large-N gauge theories contain states like $\cG_J$ for $J \geq 2$, and hence $N$ cannot be parametrically larger than $\sqrt{c_T}$.  On the AdS side, this implies that hadrons with spin $J>2$ violate causality when we take $N\rightarrow \infty$ first and then $G_N\rightarrow 0$.

\subsection{Restoring Causality and Parametric Bounds}
One way causality can be restored is by tuning $N$ such that $1/N^2$ effects can compete with the $1/c_T$ effects. In other words, our previous argument will break down if contributions from the third line of (\ref{defF}) are comparable to the contribution from stress tensor exchange. 

Before we proceed, let us note that mass of glueball states are $m_G\sim \Lambda_{\text{QCD}}$ and so if $R_{\text{AdS}} \Lambda_{\text{QCD}} \gg 1$ then the dimension of the dual operator $\cG_J$ is large: $\Delta_J \gg 1$. The contribution from the stress tensor exchange can  be schematically written as \cite{Afkhami-Jeddi:2018apj}
\be
\delta F_{\text{smeared}}|_T\sim i \frac{\Delta_0 \Delta_J}{c_T} \frac{1}{\sigma} f(\rho)
\ee
where, $f(\rho)$ is a function of $\rho$ and the polarization of the operator $\cG_J$. The factor of $\Delta_J$ comes from the three-point function $\langle \cG_J \cG_J T\rangle$. The Ward identity requires that at least one of the OPE coefficients of $\langle \cG_J \cG_J T\rangle$ must grow with $\Delta_J$. Similarly, the factor of $\Delta_0$ comes from the OPE coefficient of $\langle \cG_0 \cG_0 T\rangle$. 

As we mentioned earlier $\delta F_{\text{smeared}}|_T$ violates causality. However, now we should consider other exchanged conformal blocks. For example, exchange of a spin $s$ primary $\cG_s$ and its descendants contribute
\be
\delta F_{\text{smeared}}|_s\sim i \frac{1}{N^2} \frac{1}{\sigma^{s-1}} f_s(\rho)\ .
\ee
Individually, all of these contributions violate causality for $s>2$. So, these exchanges can only make the correlator causal if the spectrum of operators and their OPEs are highly fine tuned. In particular, the sum $\sum_s \delta F_{\text{smeared}}|_s$ should not grow faster than $1/\sigma$. Furthermore, the sum should cancel the causality violating contributions from the stress tensor. However, for that to occur, it is necessary that $1/N^2$ is not too small
\be
\frac{\Delta_0 \Delta_J}{c_T}  \lesssim \frac{1}{N^2} \ ,
\ee
where we are assuming that the sum over $s$ does not scale with $N$. The above relation can be rewritten as a bound on $N$ 
\be
N\lesssim \frac{\sqrt{c_T}}{\sqrt{\Delta_0 \Delta_J}}\ .
\ee
Note that  we find the strongest bound by considering the heaviest glueball states where the theory is under control, which would mean setting $\Delta_0, \Delta_J \sim \Delta_{\text{gap}}$.  However, using states much heavier than $\Lambda_{\text{QCD}}$ could introduce large, uncontrolled, theory-dependent form factors which could greatly alter our analysis.  Thus to be conservative, we choose $\Delta_0 \sim \Delta_J \sim R_{\text{AdS}} \Lambda_{\text{QCD}} = \Delta_{QCD}$, which leads to 
\be
N\lesssim \frac{\sqrt{c_T}}{ \Delta_{QCD}}\ .
\ee
On the gravity side, this translates into the bound 
\be\label{bound2}
N \lesssim \frac{M_{Pl}}{\Lambda_{\text{QCD}}}\ .
\ee
 It's worth noting that for causality to be restored, there must be a seemingly fine-tuned connection between these $1/N^2$ corrections and the graviton-exchange effects, which is in itself surprising.  
 
Before we conclude this section, we should clarify one thing.  $\cG_J$ is a heavy operator. For heavy operators, the smearing in (\ref{smeared}) is not required to project out $[\cG_J \cG_J]$ exchanges. However, the smearing procedure serves one other important function. It leads to optimal bounds on CFT three-point functions from causality. This property of the smearing procedure was first shown in \cite{Hartman:2016lgu}.  This fact was later used in the context of holographic CFTs in \cite{Afkhami-Jeddi:2016ntf,Afkhami-Jeddi:2018apj,Afkhami-Jeddi:2017rmx,Afkhami-Jeddi:2018own} to derive optimal constraints.

It is natural to extend this argument to mesons. Most of the analysis is exactly the same. Causality can be restored by exchanging a tower of mesonic operators $\Pi_s$  
\be
\delta F_{\text{smeared}}|_\Pi\sim \sum_s i \frac{1}{N} \frac{1}{\sigma^{s-1}} f_s(\rho)\ 
\ee
 which leads to a parametric bound
\be\label{bound3}
\sqrt{N} \lesssim \frac{M_{Pl}}{\Lambda_{\text{QCD}}}\ .
\ee
The bound from glueballs is both more general and stronger.  Unfortunately, we cannot determine the order one factors in these bounds without more precise information about the QCD sector and its form factors.

Both (\ref{bound2}) and (\ref{bound3}) are independent of the AdS radius $R_{\text{AdS}}$.   However, we have implicitly assumed that we do not need to include operator mixing effects associated with particle decay in AdS, which we discussed in section \ref{sec:OperatorMixingandUnstableParticles}.  We would expect (conservatively) that these effects could alter our analysis unless $N \gg \Lambda_{\text{QCD}} R_{\text{AdS}}$, and so we cannot apply our bounds in the flat space limit $R_{\text{AdS}} \to \infty$.  However, in subsequent sections we will take a different approach by directly studying causality bounds on flat space scattering.

\section{Eikonal Scattering in Large $N$ Gauge Theory}
\label{sec:ScatteringAnalysis}

Causality  constrains   eikonal scattering amplitudes, and these may be used to obtain a bound relating $N$ and $G_N$ in flat spacetime. It was shown in \cite{Camanho:2014apa} that the eikonal phase shift determines the Shapiro time delay and hence should be positive. The same argument was used in \cite{Afkhami-Jeddi:2018apj} to rule out massive elementary particles with spin $J>2$. In this section, we will argue that the proof of \cite{Afkhami-Jeddi:2018apj} also holds for eikonal scattering of a spin zero glueball with a higher spin glueball in certain limits, implying the bound (\ref{bound1}) for  confining large $N$  gauge theories in flat space. Bounds could also be obtained  from the  scattering of a higher spin glueball with a spectator field or graviton.

We will also explain why our bounds do not apply when they  should not.  For instance, our bounds do not constrain Kerr black holes because of finite size effects. And they do not constrain hydrogen atoms because the Bose-enhancement trick \cite{Camanho:2014apa} cannot be applied, as a high-energy scattering process with a hydrogen atom will be overwhelmingly  likely to shatter it into its constituents.  

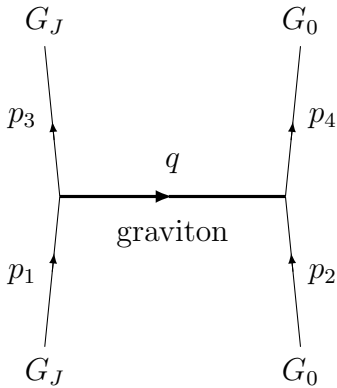
\begin{figure}
\begin{center}
\usetikzlibrary{decorations.markings}    
\usetikzlibrary{decorations.markings}    
\begin{tikzpicture}[baseline=-3pt,scale=0.50]
\begin{scope}[very thick,shift={(4,0)}]
\draw[thin,-latex]  (-3,0) -- (-3.2,2.0);
\draw[thin]  (-3.15,1.5) -- (-3.4,4.0);
\draw[thin]  (-3.2,-2.0) -- (-3,0);
\draw[thin, -latex]  (-3.4,-4)--(-3.15,-1.5) ;
\draw[thin, -latex]  (3,0) -- (3.2,2.0);
\draw[thin]  (3.15,1.5) -- (3.4,4.0);
\draw[thin]  (3.2,-2.0) -- (3,0);
\draw[thin, -latex]  (3.4,-4)--(3.15,-1.5) ;
\draw[very thick, -latex]  (-3,0)--(0,0) ;
\draw[very thick ]  (-0.1,0)--(3,0) ;
\draw(-4,1.5)node[above]{$p_3$};
\draw(-4,-1.5)node[below]{$p_1$};
\draw(4,1.5)node[above]{$p_4$};
\draw(4,-1.5)node[below]{$p_2$};
\draw(0,0.3)node[above]{ $q$};
\draw(0,-0.3)node[below]{ graviton};
\draw(-3.4,-4)node[below]{ $G_J$};
\draw(-3.4,4)node[above]{ $G_J$};
\draw(3.4,-4)node[below]{ $G_0$};
\draw(3.4,4)node[above]{ $G_0$};
\end{scope}
\end{tikzpicture}
\end{center}
\caption{\label{fig_eik} \small Eikonal scattering of glueballs in large $N$ gauge theory. In the limit $N\rightarrow \infty$, the leading non-trivial contribution comes from a graviton exchange.}
\end{figure}

\subsection{Eikonal Scattering and Causality}
Let us now summarize the main argument that we will use to derive (\ref{bound1}). First, we extend the argument of \cite{Camanho:2014apa} to constrain eikonal scattering of glueballs in confining large $N$ gauge theories. In particular, we study $2 \to 2$ scattering of a spin zero glueball  $G_0$ with a higher spin glueball $G_J$ with $J>2$, as shown in figure \ref{fig_eik}, in $(3+1)-$dimensional flat spacetime.\footnote{Let us note that it is not essential to study eikonal scattering of glueballs to derive the bound. In fact, one can replace the spin-zero glueball $G_0$ by a graviton and study the scattering: graviton+$G_J \rightarrow$ graviton+$G_J$. The argument is almost identical, however, the eikonal scattering with a graviton has one clear advantage. In this setup, it is easier to estimate the contributions to the phase shift from other exchanges (or loops) to show that they are indeed negligible in the eikonal limit.} We will use the following null coordinates in $\mathbb{R}^{1, 3}$ 
\ba
ds^2  = -du dv + d\vec{x}_{\perp}^2\ .
\ea
In the eikonal limit, both particles are highly boosted such that they are moving almost in the null directions. In other words, we are in the regime: $s\gg |t|,m_0,m_J$, where the Mandelstam variables are 
\ba
s= - (p_1+p_2)^2\ ,  \qquad t= -(p_1 -p_3)^2 = -q^2\approx -\vec{q}^{\ 2}\ .
\ea 
The masses of the glueball states are $m_0$ and $m_J$ respectively. 

The tree-level scattering amplitude, when expressed in the impact parameter space $\vec{b}$, is known as the phase shift:
\ba\label{phaseshift}
 \delta (s,\vec{b}) = \frac{1}{2 s} \int \frac{d^{2}\vec{q}}{(2 \pi)^{2}} \; e^{i \vec{q} \cdot \vec{b}} M_{\text{tree}}(s, {\vec{q}\;})\ .
\ea
It is expected that only ladder diagrams contribute in the eikonal limit and hence the total amplitude is given by the exponential of the tree level phase shift. This phase shift, when exponentiated, can be interpreted as the Shapiro time-delay experienced by either of the particles and hence must be non-negative \cite{Camanho:2014apa} (also see section 2 of  \cite{Afkhami-Jeddi:2018apj})
\be
\delta (s,\vec{b}) \ge 0\ .
\ee
However, it is not completely obvious that the tree-level amplitude must exponentiate in the eikonal limit. Furthermore, glueballs/mesons have finite size and hence the statement of causality should be modified as well. 
 
\subsection{Causality Condition for a Particle with a Finite Size}
Let us now revisit the $\N$-shockwave setup of \cite{Camanho:2014apa} but for particles with finite size. This setup has several advantages, for example in this setup the phase shift (\ref{phaseshift}) naturally exponentiates.

\begin{figure}
\centering
\includegraphics[scale=0.31]{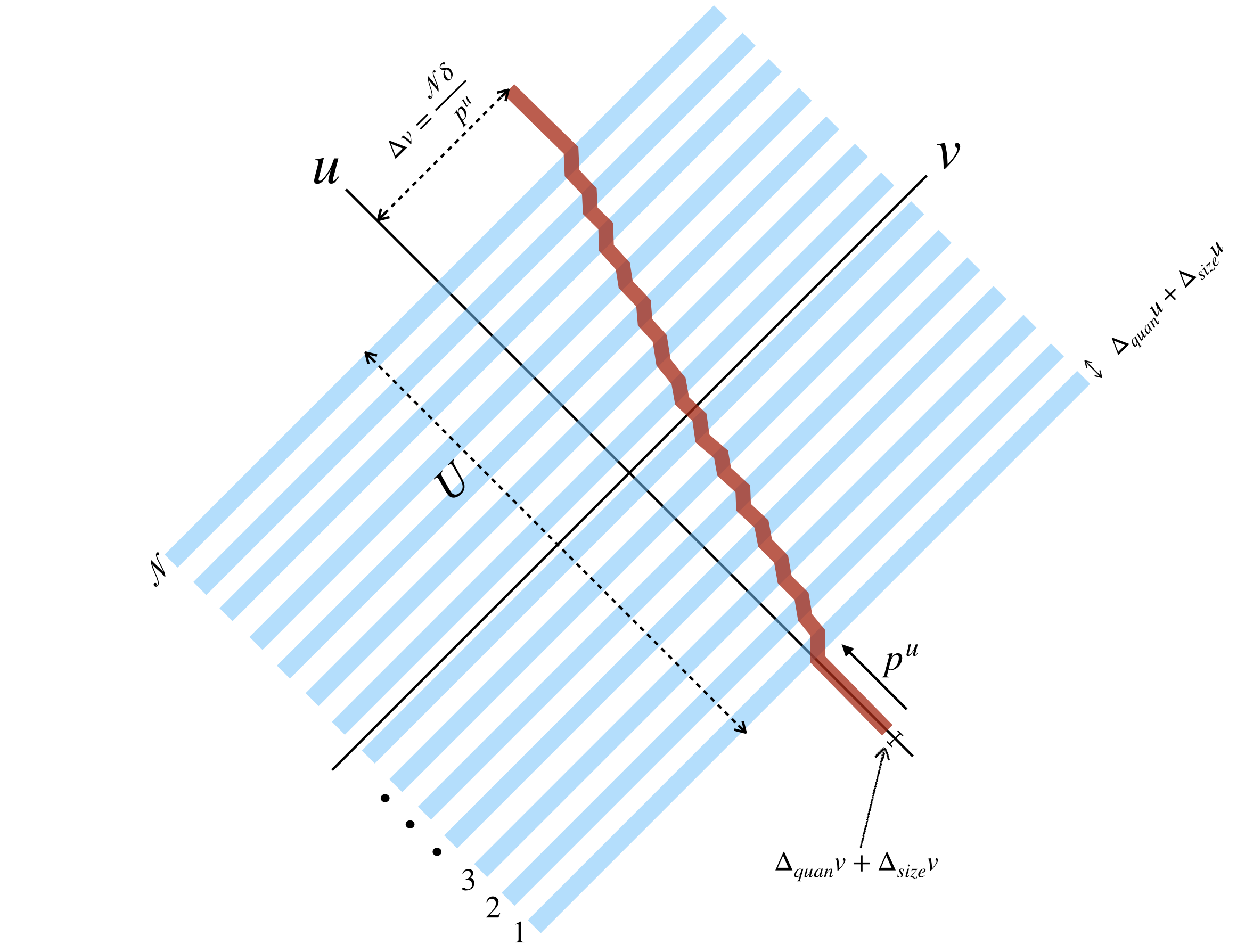}
\caption{\small Total time delay for a coherent state of incoming particles after crossing $\N$ independent shockwaves can be large enough to violate asymptotic causality. }\label{nshock}
\end{figure}

First, let us note that the eikonal scattering can be thought of as the particle $G_J$ traveling in a shockwave sourced by the other particle $G_0$. At tree-level, the amplitude is $1+i \delta$, where $\delta \ll 1$ in order for the theory to be weakly coupled. The Shapiro time delay of the particle $G_J$ is related to the phase shift  in the following way
\be
\Delta v= \frac{\delta}{p^u}\ ,
\ee
where, $p^u>0$ is the $u$-component of the momentum of particle $G_J$ (see figure \ref{nshock}). Naively, one would expect that causality requires $\delta  \ge 0$. However, that is not exactly correct. In order for a time advance to imply causality violation, $|\Delta v|$ should be larger than all uncertainties associated with the thought experiment \cite{Camanho:2014apa}. There are two types of uncertainties: (i) the quantum mechanical uncertainty of the wave packet
\be
\Delta_{\text{quan}} v \sim \frac{1}{p^u}\ ,
\ee
and (ii) uncertainty due to the finite size of the particle
\be
\Delta_{\text{size}} v \sim \frac{m_J r}{p^u}\ ,
\ee
where, $r$ is the size of the particle with mass $m_J$. The particle $G_J$ is highly boosted and the factor of $m_J/p^u$ takes into account the requisite length contraction. 

The finite size effect is important because in order for us to detect a time advance, we first need to find the lightcone associated with the initial state. $\Delta_{\text{size}} v$ can be thought of as a source of error in determining the lightcone of the particle. If a particle of mass $m$ satisfies $m  r \ll 1$, we can consider the particle effectively elementary. However, for mesons or guleballs  $m r\gtrsim 1$ and hence  the finite size effect is more important than the quantum effect. So, to be conservative, for glueballs a causality violation can only be detected if
\be
|\delta| > m_J r\ .
\ee

Perturbation theory requires that $|\delta| \ll 1$. This small perturbative effect can be amplified by studying the propagation of the particle $G_J$ in a background with $\mathcal{N}$ independent shockwaves  created by $G_0$ particles \cite{Camanho:2014apa} (for a pictorial representation see  figure \ref{nshock}). For amplification, it is required that the phase shift $\delta$ is the same for each of these $\N$-processes. This happens naturally if the polarization of the outgoing $G_J$ is  the complex conjugate of that of the incoming $G_J$. This post selection can be achieved through Bose enhancement -- replacing the incoming particle $G_J$ by a coherent state of particles with a fixed polarization \cite{Camanho:2014apa}. Since mesons and glueballs are weakly interacting in the large $N$ limit, we can tune the mean occupation number to be large but still have small $\delta$ at each step. Then Bose enhancement ensures that the incoming and outgoing states are exactly the same. Of course, when $N$ is large but finite the process of post selection by Bose enhancement is more subtle;  we will address this at the end of this section.

In the limit $\delta \rightarrow 0$ and $\mathcal{N}\rightarrow \infty$ with $\mathcal{N}\delta$ fixed, the total amplitude is 
\be
(1+i \delta)^\mathcal{N}\approx e^{i \mathcal{N} \delta}\ .
\ee 
Therefore, the total phase-shift is $\mathcal{N} \delta$.  The absolute value of this quantity should be larger than $m_J r$ if we are to definitively observe a time advance. 

Next, we also need to make sure that the center of the particle $G_J$ remains localized on the transverse plane at a distance $b$ through the entire process. Let us assume that the entire process of scattering through $\mathcal{N}$ shocks takes null time $U$. In that null time, the wavefunction for the center of the particle $G_J$ spreads along the transverse direction by \cite{Camanho:2014apa}
\be
\Delta b \sim \sqrt{\frac{U}{p^u}}\ .
\ee
The particle $G_0$ (or rather a coherent state of particle $G_0$) with momentum $P^v$ that creates a shock can only be localized over a distance\footnote{For now let us assume that the particle $G_0$ has zero size for simplicity. It is straightforward to include the finite size effect of $G_0$ by considering
\be
\Delta_{\text{size}} u \sim \frac{m_0 r'}{P^v}\ ,
\ee
where, $m_0$ is the mass and $r'$ is the size of the particle $G_0$. 
} 
\be
\Delta_{\text{quan}} u \sim \frac{1}{P^v}\ .
\ee
Therefore, we can only get $\mathcal{N}$ independent shocks if 
\be
U=\frac{\mathcal{N}}{P^v}\ 
\ee
implying 
\be
\Delta b \sim \sqrt{\frac{\mathcal{N}}{s}}\ .
\ee
Therefore, the particle $G_J$ can only be localized at a distance $b$, if
\be
\mathcal{N}\ll b^2 s\ .
\ee
Moreover, using $\mathcal{N}\sim m_J r/|\delta|$, we obtain
\be\label{grrm}
\frac{m_J r}{b^2 s}\ll |\delta| \ll 1\ .
\ee
Note that causality violation, if any, shows up in the phase-shift only in the regime $b\lesssim 1/m_J$. So, taking $s\sim \Lambda_{\text{UV}}^2$ and $b\sim 1/m_J$ we get a bound on the size of the particle which can be constrained using eikonal scattering
\be\label{finite}
\left(\frac{m_J}{\Lambda_{\text{UV}}} \right)^2 m_J  r \ll 1\ .
\ee
Both mesons and glueballs have $m\sim \Lambda_{\text{QCD}}$ and $m r \sim 1$ and hence they obey this condition. 
Note that a macroscopic black hole \emph{would not} obey this condition, and hence Kerr black holes with spin more than two are not ruled out by this argument. 

Finally, let us consider an eikonal scattering where particle $G_0$ also has a finite size $r'$ with $m_0 r'>1$. In that case, 
\be
U=\frac{\mathcal{N} m_0 r'}{P^v}
\ee
and hence the condition (\ref{grrm}) now becomes
\be
\frac{(m_J r)(m_0 r')}{b^2 s}\ll |\delta| \ll 1\ .
\ee

Mesons and glueballs are stable particles strictly in the $N\rightarrow \infty$ limit. For, large but finite $N$, both mesons and glueballs do decay. The lifetime of a meson is $\sim \O(N)$. Whereas, typical lifetime of a glueball is $\O(N^2)$. However, these decay processes do not immediately invalidate our argument. If we perform our experiment with an ensemble of incoming particles, the finite lifetime implies that only a fraction of these incoming particles can experience a Shapiro time delay/advance. Detection of time advance even for a single particle is sufficient to conclude that the theory is acausal.

The discussion of section \ref{hydrogen} implies that the condition (\ref{finite}) is just a necessary condition but not sufficient. For example, naively the condition (\ref{finite}) suggests that hydrogen atoms can be treated as elementary particles in the regime of interest. However, hydrogen atoms with spin more than two are not ruled out  because in gravitational eikonal scattering hydrogen atoms cannot be approximated as elementary particles. In fact, the leading contribution to the hydrogen phase shift comes from the protons and the electron interacting with the graviton individually. The weak electromagnetic interaction between the electron and the proton can only contribute to the phase shift at the subleading order.  Whereas, confinement ensures that glueballs and mesons behave as elementary particles in the large $N$ limit.

\subsection{Bose Enhancement and Mesons/Glueballs Mixing}

For glueball or meson scattering there can be mixing between different states when a high energy graviton hits a glueball/meson. One might expect that this kind of mixing should not contribute to our causality argument because of the post selection through Bose enhancement. Let us now make that expectation more precise by studying when the Bose enhancement argument of \cite{Camanho:2014apa} can breaks down.  We will see that for dominantly inelastic scattering processes,  such as high energy scattering of Hydrogen atoms, our bounds do not apply.

\subsubsection{Bose Enhancement}
Let us consider the operator $a^\dagger$ that creates a single particle state (elementary or composite) with a particular polarization. For our flat space positivity argument it is essential that the phase shift exponentiates. This was achieved by  considering $\mathcal{N}$-shockwaves following  \cite{Camanho:2014apa}. Moreover, our argument requires that the higher spin particle (glueball state) should be replaced by a coherent state of the same higher spin particles (glueballs)
\be
|i\rangle = e^{\lambda a^\dagger}|0\rangle\ ,
\ee
where, $\lambda$ is real. Before we introduce interactions, note that $\langle i' | i\rangle$ where  $|i'\rangle=e^{\lambda' a^\dagger}|0\rangle$:
\be\label{overlap}
\langle i' | i\rangle= e^{\lambda \lambda'}\ .
\ee
We now introduce the following interaction
\be
H_{int}=\delta_1  a^\dagger a + \delta_2  b^\dagger a + \delta_2^* a^\dagger b
\ee
where, $\delta_1$ is the phase shift for the process: state $a$ goes to state $a$. Similarly $\delta_2$ is the phase shift for the process: state $a$ goes to state $b$.

We can now consider a basis of outgoing states:
\begin{align}
|\lambda';0\rangle=e^{\lambda' a^\dagger}|0\rangle\ ,\quad |\lambda';1\rangle=b^\dagger e^{\lambda' a^\dagger}|0\rangle\ ,\quad |\lambda';2\rangle=(b^\dagger)^2e^{\lambda' a^\dagger}|0\rangle\ ,\quad  \cdots \ .
\end{align}
Matrix elements can be obtained by using equation (\ref{overlap}), yielding (after properly normalizing all states)
\begin{align}
\langle  \lambda;0| H_{int}|i\rangle = \lambda^2 \delta_1\ , \qquad \langle  \lambda;1| H_{int}|i\rangle = \lambda \delta_2\ 
\end{align}
and all other matrix elements are zero.

\subsubsection{A Condition for Bose Enhancement} 

From the above matrix elements it appears  like $|i\rangle \rightarrow |i\rangle$ is enhanced when $\lambda\gg 1$. However, one should be more careful when there is a parametric separation between $\delta_1$ and $\delta_2$. First of all, the theory is weakly coupled only when 
\be
|\delta_1| \ll 1\ , \qquad |\delta_2| \ll 1\ .
\ee
For an enhancement of the process $|i\rangle \rightarrow |i\rangle$, we require that  the mean occupation number should be large enough so that we can neglect the second process
\be
\lambda^2 |\delta_1| \gg \lambda |\delta_2|\ .
\ee
However, the mean occupation number should also be small enough that the total scattering amplitude is still small
\be
\lambda^2 |\delta_1| \ll 1\ , \qquad \lambda |\delta_2|\ll 1\ .
\ee
All these conditions can only be satisfied if
\be\label{BEcon}
|\delta_1| \gg |\delta_2|^2\ .
\ee
Therefore, the Bose enhancement argument cannot be trusted if $\delta_2^2 \sim \delta_1$.  

\subsubsection{Failure of Bose Enhancement for Hydrogen Atoms}

There is one more possibility that we must consider -- the final state $b$ can be highly degenerate.  Physically, this includes scenarios where the scattering process is dominantly inelastic, and the interaction may shatter the initial state into a large number of final states.  

We can parameterize the Hamiltonian as
\be
H_{int}=\delta_1  a^\dagger a + \sum_{i=1}^{n_{deg}} \left[ \delta_2  b_i^\dagger a + \delta_2^* a^\dagger b_i \right]
\ee
where generally $\delta_2$ will depend on the final state, but we have suppressed this for simplicity.
We can Bose-enhance the process $a\rightarrow a$ if 
\be
\lambda^2 |\delta_1| \gg \lambda n_{deg} |\delta_2|\ ,
\ee
This can only be achieved without going beyond the weakly coupled regime if
\be
|\delta_1| \gg n_{deg}^2 |\delta_2|^2\ .
\ee
Therefore, the Bose enhancement trick will fail if the degeneracy of the other states $b_i$ is very large, in particular it fails once
\be\label{BEcon2}
n_{deg} \gtrsim \frac{\sqrt{\delta_1}}{\delta_2}\ .
\ee
This condition explain why the causality bound cannot be applied to hydrogen atoms with spin $J>2$. The derivation of the causality bound heavily relies on the post selection of the final state with the help of Bose enhancement. But in an eikonal scattering of hydrogen atoms,  when the energy of the exchanged graviton is large compared to the mass of a hydrogen atom, the scattering process will almost always shatter the atom into a proton, an electron, and many photons.  The condition (\ref{BEcon2}) immediately implies that the Bose enhancement trick breaks down for high energy eikonal scattering of hydrogen atoms. 

This reasoning does not invalidate the bound for hadrons in confining large $N$ gauge theories, because the processes that shatter the hadrons are suppressed by additional powers of $1/N$ as compared to scattering that preserves the initial hadron, as shown for the scalings in equation (\ref{eq:GGTScaling}).

\subsubsection{Mixing}

We want to enhance the process of figure \ref{fig_eik} where the polarization $\epsilon_3$ of the outgoing $G_J$ is the complex conjugate of the polarization $\epsilon_1$ of the incoming $G_J$. This can be done using Bose enhancement because the phase shift for $\epsilon_3=\epsilon_1^*$ is of the same order as the phase shift for $\epsilon_3\neq \epsilon_1^*$ and the condition (\ref{BEcon}) is trivially satisfied. 

The same is true even for mixing between different glueball (or meson) states. Since glueballs and mesons are composite particles, a high energy graviton can change the internal state of a glueball (or meson) by converting it into a  different hadron which may have different mass and/or spin. This type of mixing is not suppressed by the large $N$ limit (see equation \ref{eq:GGTScaling}) and hence a priori these mixings should not be ignored for glueball/meson eikonal scattering.  

If we start with an incoming glueball state $G_J$, there are two things that can happen as shown below 
\be\label{be}
\begin{gathered}\includegraphics[width=0.79\textwidth]{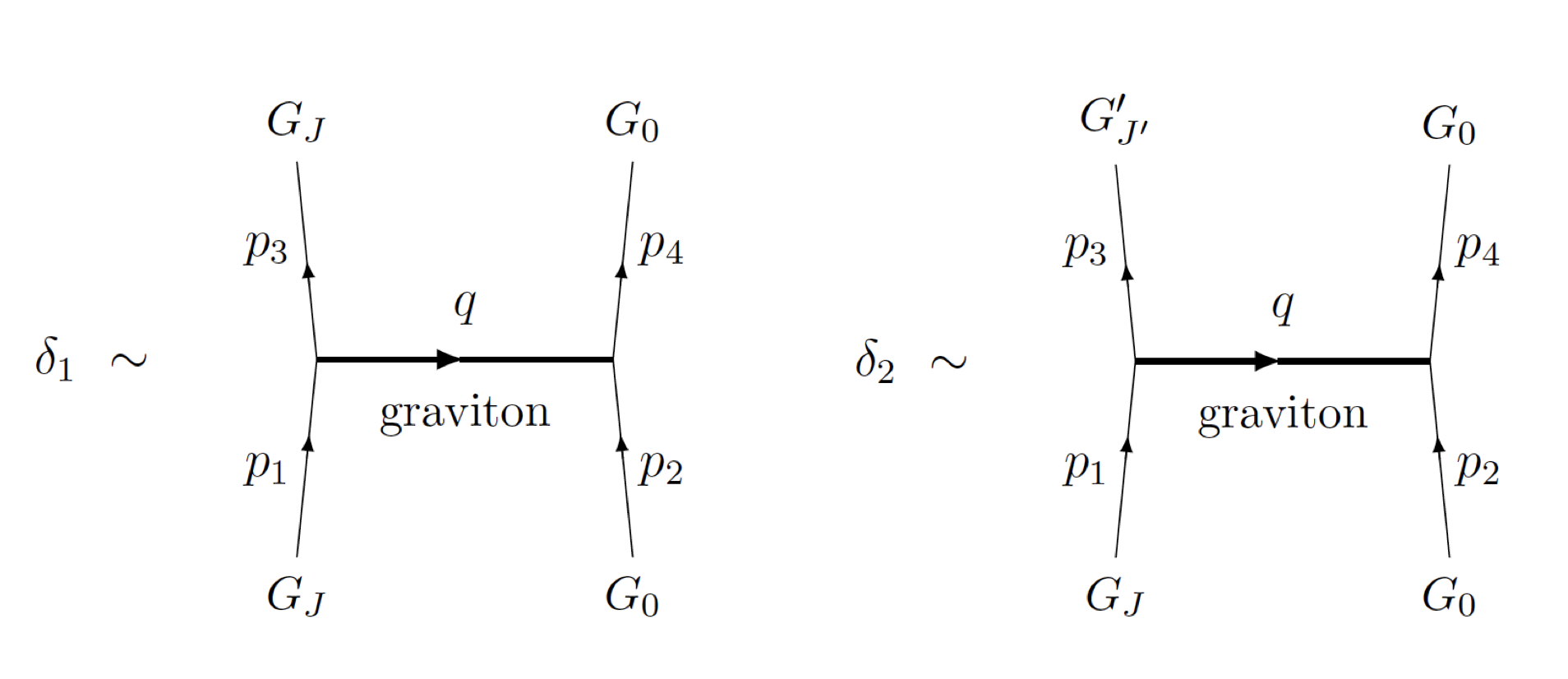}\end{gathered}\ ,
\ee
where, for simplicity we only consider mixing of the spinning glueball $G_J$ but not $G_0$. We can estimate $\delta_1$ and $\delta_2$. In the eikonal limit, the behavior of the phase shift is completely fixed by the exchanged particle and hence $\delta_1, \delta_2\sim s/M_{Pl}^2$. Thus, we can project to the first process by Bose enhancement.

\subsection{A Bound on $N$}\label{boundN}
The proof of \cite{Afkhami-Jeddi:2018apj} holds for eikonal scattering of a spin zero glueball with a higher spin glueball in a confining  large $N$ gauge theory, however, there are several subtleties as we explained before. There is an additional technical subtlety exclusively in $(3+1)$-dimensional spacetime because of the glueball mixing which we will address next.

\subsubsection*{Tree-Level Eikonal Scattering}

Let us now come back to the tree-level eikonal scattering amplitude of a higher spin glueball $G_J$ with a scalar glueball $G_0$ for a large $N$ gauge theory, as shown in figure \ref{fig_eik}.\footnote{There is nothing special about $G_0$. For example, one can replace the glueball $G_0$ by a graviton and make  the same argument. } The Bose enhancement trick implies that the polarization $\epsilon_3$ of the outgoing $G_J$ is the complex conjugate of the polarization $\epsilon_1$ of the incoming $G_J$ and we can ignore mixing.  If $N$ is the largest quantity of the theory, then we  only need to consider graviton exchanges and the phase shift is completely fixed by the on-shell three-point amplitude $G_J G_J h_{\mu \nu}$, where $h_{\mu\nu}$ is the graviton. In general the on-shell three-point amplitude $G_J G_J h_{\mu \nu}$ can be any linear combination of $2J+1$ parity even and $2J$ parity odd structures (see section 2.5 of \cite{Afkhami-Jeddi:2018apj}). The phase shift is schematically given by
\be
\delta (s,\vec{b}) \sim \frac{s}{M_{Pl}^2} f\left(\frac{\vec{\partial}_b}{m_J}\right) \ln\left( \frac{L}{b}\right)\label{gr}
\ee
where, $L$ is the IR regulator and $f$ is some known function described in  \cite{Afkhami-Jeddi:2018apj}. Causality requires that this phase shift should be positive for any polarization of the incoming particle $G_J$. As shown in \cite{Afkhami-Jeddi:2018apj}, this phase shift in the limit $m_J b \ll 1$ violates causality for all glueballs with spin $J>2$ unless   the on-shell three-point amplitude $G_J G_J h_{\mu \nu}$ is a very specific combination of only parity even structures. This specific combination corresponds to a non-minimal coupling between glueballs and gravitons. The same conclusion holds for higher spin mesons as well.

This remaining non-minimal coupling can be ruled out by applying interference bounds,  as shown in \cite{Afkhami-Jeddi:2018apj}. However, for composite particles such as glueballs we need to be more careful because on-shell three-point amplitudes of mixing $G'_{J'} G_J h_{\mu \nu}$ can contribute significantly in the interference setup. So, first we need to bound the mixing amplitude  $G'_{J'} G_J h_{\mu \nu}$ from causality.

\subsubsection*{A Bound on On-Shell Mixing Amplitudes}

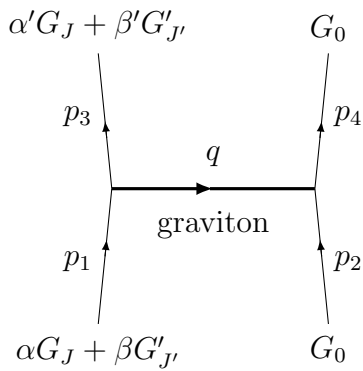
\begin{figure}
\begin{center}
\usetikzlibrary{decorations.markings}    
\usetikzlibrary{decorations.markings}    
\begin{tikzpicture}[baseline=-3pt,scale=0.45]
\begin{scope}[very thick,shift={(4,0)}]
\draw[thin,-latex]  (-3,0) -- (-3.2,2.0);
\draw[thin]  (-3.15,1.5) -- (-3.4,4.0);
\draw[thin]  (-3.2,-2.0) -- (-3,0);
\draw[thin, -latex]  (-3.4,-4)--(-3.15,-1.5) ;
\draw[thin, -latex]  (3,0) -- (3.2,2.0);
\draw[thin]  (3.15,1.5) -- (3.4,4.0);
\draw[thin]  (3.2,-2.0) -- (3,0);
\draw[thin, -latex]  (3.4,-4)--(3.15,-1.5) ;
\draw[very thick, -latex]  (-3,0)--(0,0) ;
\draw[very thick ]  (-0.1,0)--(3,0) ;
\draw(-4,1.5)node[above]{$p_3$};
\draw(-4,-1.5)node[below]{$p_1$};
\draw(4,1.5)node[above]{$p_4$};
\draw(4,-1.5)node[below]{$p_2$};
\draw(0,0.3)node[above]{ $q$};
\draw(0,-0.3)node[below]{ graviton};
\draw(-3.4,-4)node[below]{ $\alpha G_J+ \beta G'_{J'}$};
\draw(-3.4,4)node[above]{ $\alpha' G_J+ \beta' G'_{J'}$};
\draw(3.4,-4)node[below]{ $G_0$};
\draw(3.4,4)node[above]{ $G_0$};
\end{scope}
\end{tikzpicture}
\end{center}
\caption{\label{fig_mix} \small A setup to bound glueball mixing  in large $N$ gauge theories. In the limit $N\rightarrow \infty$, the leading non-trivial contribution still comes from a graviton exchange.}
\end{figure}
We now consider an eikonal scattering: $1,2\rightarrow 3,4$, where, 1 and 3 are linear combinations $\alpha G_J+ \beta G'_{J'}$ and $\alpha' G_J+ \beta' G'_{J'}$ respectively with real coefficients $\alpha,\alpha',\beta,\beta'$ (see figure \ref{fig_mix}). Particles 2 and 4 are again a scalar glueball $G_0$. Causality now  can be expressed as semi-definiteness of the phase shift matrix $\delta_{13}$:
\begin{align}\label{matrix}
\delta_{13}\equiv \left(
\begin{array}{cc}
\delta_{GG} & \delta_{GG'}\\ 
\delta_{GG'}^* & \delta_{G'G'}
\end{array} 
\right)\succeq 0\ .
\end{align}
The above condition can also be restated as a bound on $\delta_{GG'}$:
\be\label{intf_bound}
|\delta_{GG'}|^2 \le \delta_{GG}\delta_{GG'}\ .
\ee
Positivity of $\delta_{GG}$ and $\delta_{GG'}$ for all polarizations implies that \cite{Afkhami-Jeddi:2018apj}
\be
\delta_{GG}= a_1 \frac{s}{M_{Pl}^2} \ln \left(\frac{L}{b} \right)\ , \qquad \delta_{G'G'}= a_1' \frac{s}{M_{Pl}^2} \ln \left(\frac{L}{b} \right)\ ,
\ee
where, $a_1$ and $a_1'$ are dimensionless coefficients. Hence, $\delta_{GG'}$ should not grow faster than $\frac{s}{M_{Pl}^2} \ln \left(\frac{L}{b} \right)$ in the limit $\Lambda_{\text{UV}}\gg 1/b\gg m_G, m_{G'}$.

One immediate consequence of the above growth bound is that in the limit $N\rightarrow \infty$, the on-shell three-point amplitude (see appendix \ref{mixing} for details)
\be\label{bound_mixing}
G'_{J'} G_J h_{\mu \nu} \lesssim \frac{1}{M_{Pl}} \frac{\ln(\Lambda_{\text{UV}} L)}{\Lambda_{\text{UV}}^{n}}\qquad \text{with} \qquad n\ge 1 \ .
\ee
Since $GG h_{\mu \nu} \sim 1/M_{Pl}$, the glueball mixing $GG' h_{\mu\nu}$ is always suppressed by the UV cut-off scale. The same conclusion holds even for the meson mixing $\pi \pi' h_{\mu\nu}$. 

\subsubsection*{Graviton Interference Bound}

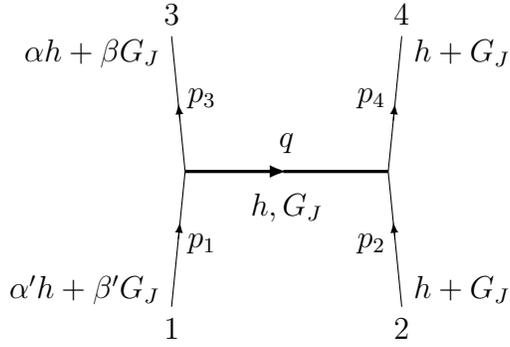
\begin{figure}
\begin{center}
\usetikzlibrary{decorations.markings}    
\usetikzlibrary{decorations.markings}    
\begin{tikzpicture}[baseline=-3pt,scale=0.45]
\begin{scope}[very thick,shift={(4,0)}]

\draw[thin,-latex]  (-3,0) -- (-3.2,2.0);
\draw[thin]  (-3.15,1.5) -- (-3.4,4.0);
\draw[thin]  (-3.2,-2.0) -- (-3,0);
\draw[thin, -latex]  (-3.4,-4)--(-3.15,-1.5) ;

\draw[thin, -latex]  (3,0) -- (3.2,2.0);
\draw[thin]  (3.15,1.5) -- (3.4,4.0);
\draw[thin]  (3.2,-2.0) -- (3,0);
\draw[thin, -latex]  (3.4,-4)--(3.15,-1.5) ;

\draw[very thick, -latex]  (-3,0)--(0,0) ;
\draw[very thick ]  (-0.1,0)--(3,0) ;

\draw(-2.5,1.5)node[above]{$p_3$};
\draw(-2.5,-1.5)node[below]{$p_1$};
\draw(2.5,1.5)node[above]{$p_4$};
\draw(2.5,-1.5)node[below]{$p_2$};
\draw(0,0.2)node[above]{ $q$};
\draw(0,-0.3)node[below]{ $h, G_J$};
\draw(-3.4,-4)node[below]{ $1$};
\draw(-3.4,4)node[above]{ $3$};
\draw(3.4,-4)node[below]{ $2$};
\draw(3.4,4)node[above]{ $4$};

\draw(3.4,3.5)node[right]{ $h+G_J$};
\draw(3.4,-3.5)node[right]{ $h+G_J$};
\draw(-3.4,3.5)node[left]{ $\alpha h +\beta G_J$};
\draw(-3.4,-3.5)node[left]{ $\alpha' h +\beta' G_J$};
\end{scope}
\end{tikzpicture}
\end{center}
\caption{\label{fig_int} \small Graviton interference bound: in-states are linear combinations of  $G_J$ and the graviton $h$. }
\end{figure}

Let us now study the eikonal scattering shown in figure \ref{fig_int} -- states 1 and 3 are linear combinations of $G_J$ and the graviton: $\alpha h +\beta G_J$ and $\alpha' h+ \beta' G_J$ respectively, where $\alpha,\alpha',\beta,\beta'$ are some arbitrary real coefficients. States 2 and 4 are a fixed combination of $G_J$ and the graviton: $ h + G_J$. Suppression of glueball mixing $GG'h_{\mu\nu}$ ensures that only gravitons and $G_J$ can be exchanged in the limit $N\rightarrow \infty$ -- this is  exactly  the interference setup of section 2.5 of \cite{Afkhami-Jeddi:2018apj}. This implies that the interference bound of  \cite{Afkhami-Jeddi:2018apj} applies here yielding 
\be
G_J G_J h_{\mu\nu}=0 \qquad \text{for}\qquad J>2\ 
\ee
which contradicts  the equivalence principle. The same conclusion holds even for $\pi_J\pi_J h_{\mu\nu}$. Therefore, there is no consistent way of coupling higher spin glueballs/mesons with gravity in the limit $N\rightarrow \infty$.

\subsubsection*{Exchange of an Infinite Tower of Glueball States}
Confining large $N$ gauge theories can still have higher spin mesons/glueballs if large $N$ effects can compete with $M_{Pl}$. In this scenario, an infinite tower of massive higher spin glueballs will also contribute to the phase shift. The contribution of a glueball with mass $m_j$ and spin $j$ to the phase shift is given by
\be\label{above}
\delta_j \sim \frac{1}{N^2}\left(\frac{s}{\Lambda^2} \right)^{j-1} f_j\left(\frac{\vec{\partial}_b}{m_j}\right) K_{0}(m_j b)\ ,
\ee
where, $\Lambda$ is some mass scale, $K_0$ is the Bessel-$K$ function and $f_j$ is a differential operator that can be found following \cite{Afkhami-Jeddi:2018apj}. For large $j$, $\delta_j$ can be order 1 for $s> \Lambda^2$ and hence we can think of $\Lambda$ as the cut-off scale $\Lambda_{\text{UV}}$.  Also note that if $m_j\gg m_J$, then at the scale $b\sim 1/m_J$, the phase shift $\delta_j \sim e^{-m_j/m_J}$. So, in order to get a significant contribution $m_j \sim m_J$.

Even without knowing the details, we can estimate the total contributions from the exchange of  an infinite tower of glueball states. First, let us ignore gravity completely. The phase shift now is given by a sum of the above expression (\ref{above}) over all particles exchanged in the process. Since, the individual contribution violates causality, it must be an infinite sum. Moreover, the sum should not grow faster than $s$ or else the infinite sum will also violate causality \cite{Camanho:2014apa}. We also expect that this sum will cancel the causality violation of the graviton exchange after we turn on gravity. Hence, this sum must grow  so that it can compete with (\ref{gr})
\be
\sum_j \delta_j \sim \frac{1}{N^2}\left(\frac{s}{\Lambda_{\text{UV}}^2} \right)^a\ 
\ee
with $a\lesssim 1$. A comparison of the above expression with (\ref{gr}) leads to an approximate upper bound on $N$. The strongest bound is obtained at largest energies $s\sim \Lambda_{\text{UV}}^2$ which yields
\be
N \lesssim \frac{M_{Pl}}{\Lambda_{\text{UV}}}\ .
\ee
One can repeat the same argument for mesons which again leads to a weaker bound $\sqrt{N} \lesssim M_{Pl}/\Lambda_{\text{UV}}$ due to the different large $N$ scalings of meson scattering amplitudes.

\section{A Species Bound from Entropy}
\label{sec:SpeciesBoundEntropy}

Let us now present a simple entropic argument\footnote{Similar arguments and bounds are well-known, see for example \cite{Bekenstein:1980jp, Susskind:1994sm, Bousso:1999xy, Veneziano:2001ah, Dvali:2007hz}} that provides the same upper bound on $N$. Consider large $N$ QCD in $(3+1)$-dimensional flat spacetime at temperature $T$. In this thermal theory, we imagine a spherical region of radius $r$, where $r$ satisfies
\be
r=\frac{1}{4\pi T}\ .
\ee
Note that this is exactly the relation between the Hawking temperature of a spherical black hole and its radius.

A simple dimensional analysis suggests that the entropy of this spherical region is given by
\be
S\sim r^3 T^3 N^2\ .
\ee  
Finite temperature lattice computations support this expectation above the critical deconfinement temperature \cite{Panero:2009tv}. If we now increase $N$, this spherical region of radius $r$ will have the same entropy as a black hole when 
\be
S= 8\pi^2 r^2 M_{Pl}^2\ ,
\ee
where the right hand side is the Bekenstein-Hawking entropy of a Schwarzschild black hole of radius $r$. This equality holds when
\be
 \frac{N}{M_{Pl}}\sim r\ .
\ee
The large $N$ QCD plasma has more entropy than a black hole with radius $\lesssim N/M_{Pl}$. However, this may be avoided if $r$ is smaller than the UV cut-off $r<1/\Lambda_{\text{UV}}$ which translates to an upper bound on $N$:
\be
N \lesssim \frac{M_{Pl}}{\Lambda_{\text{UV}}}\ .
\ee
Since this bound is derived from UV considerations, the same is expected to be true for large $N$ QCD in AdS$_4$. 

It is  interesting that this simple and  naive argument led to the same bound on $N$. It is tempting to interpret this observation as an evidence in favor of the entropic argument (or something similar) presented in this section. However, it seems that the bound obtained in this section is stronger than that of previous sections because the entropic argument does not require any assumption about confinement or presence of stable higher spin glueballs/mesons. This suggests that a more formal argument along this line might be applicable to any large $N$ theories yielding similar bounds.

\section{Summary \& Discussion}
\label{sec:Discussion}

\subsubsection*{A Weak-Gravity Like Species Bound }
In this paper we analyzed the implications of Lorentz invariance, unitarity, and causality on large $N$ gauge theories coupled to gravity in $(3+1)$-dimensions.  We found that confining large $N$ gauge theories must obey the  species bound
\be
N \lesssim \frac{M_{Pl}}{\Lambda_{\text{QCD}}}\ , \nonumber 
\ee
though this bound is  parametric, and  the (unknown) order-one factors  will be theory-dependent. 
A simple consequence of the species bound is that the typical mass of a baryon $M \sim N \Lambda_{\text{QCD}}$ must be below the Planck scale $M_{Pl}$. 

The above species bound is precisely the weak-gravity bound for hadrons of confining large $N$ gauge theories in $(3+1)$-dimensions. The gravitational interaction between two glueballs with typical masses $\Lambda_{\text{QCD}}$ scales as $\sim \frac{\Lambda_{\text{QCD}}^2}{M_{Pl}^2}$. The species bound ensures that gravitational interaction is weaker than the gauge interaction between these glueballs which scales as $\sim 1/N^2$.

The species bound was obtained by exploring causality constraints on gravitational interactions of massive composite particles with spin $J>2$ in both AdS and flat spacetime. We showed that gravitational interactions between higher spin glueballs and mesons in any confining large $N$ gauge theory, by itself,  violate causality at $N=\infty$. Hence, a confining large $N$ gauge theory can be coupled to gravity in a consistent way  only if the gravitational interaction between hadrons is weaker than the gauge interactions between them. 

Moreover, the  eikonal scattering thought experiment in flat space as well as a rough entropic argument impose a stronger constraint which can be interpreted as a parametric bound on a UV cut-off scale of the combined gauge and gravity theory
\be
\Lambda_{\text{UV}}\lesssim \frac{M_{Pl}}{N}\ . \nonumber
\ee

\subsubsection*{Bounds on Composite Particles}

We discussed various reasons why causality bounds may not apply, including finite size effects, mixings or instabilities, and large inelastic scattering cross sections at high energy.  Weakly bound states are not constrained by causality, as high-energy scattering will simply dissociate them.  We have argued that our bounds should still apply to  high-spin hadrons with masses near the QCD scale.  But we cannot determine the order-one factors in the bounds without a much better understanding of form factors in the scattering process.  So our bounds are only parametric.

\subsubsection*{AdS vs Flat Spacetime}

In spite of  the apparent similarities between the AdS argument and the flat space argument, there are some clear differences. The AdS argument is conceptually cleaner because the statement of causality in CFT is better understood. The flat space argument appears to impose a stronger bound on $N$, though it depends on a much more intricate argument for its justification. The flat space argument also relies on the positivity of the phase-shift, which is believed to hold in all UV complete Lorentzian QFTs.  However, a rigorous S-matrix based proof is still lacking. Furthermore, the difference between the two methods becomes even more significant if we add a spectating scalar field to the confining large $N$ gauge theory. In this case, the flat-space bound on $N$ remains unchanged, whereas the CFT-based argument  leads to a stronger but theory-dependent bound. These differences, though surprising, do not  indicate any contradiction.  The operator-mixing effects in CFT associated with particle decays in AdS imply that the AdS bounds we have derived are not applicable in the flat space limit.

\subsubsection*{Universality of Gravitational Interactions of Hadrons at Large $N$}
It is important to note that the bound (\ref{bound}) is a necessary condition but not sufficient. Confining large $N$ gauge theories that obey the bound (\ref{bound}) might still violate causality. In particular, in the presence of a spectator scalar field, glueballs and mesons of all confining large $N$ gauge theories in $(3+1)$ spacetime dimensions must interact with gravity in a universal way.

\begin{figure}
\begin{center}
\usetikzlibrary{decorations.markings}    
\usetikzlibrary{decorations.markings}    
\begin{tikzpicture}[baseline=-3pt,scale=0.45]
\begin{scope}[very thick,shift={(4,0)}]
\draw[thin,-latex]  (-3,0) -- (-3.2,2.0);
\draw[thin]  (-3.15,1.5) -- (-3.4,4.0);
\draw[thin]  (-3.2,-2.0) -- (-3,0);
\draw[thin, -latex]  (-3.4,-4)--(-3.15,-1.5) ;
\draw[thin, -latex]  (3,0) -- (3.2,2.0);
\draw[thin]  (3.15,1.5) -- (3.4,4.0);
\draw[thin]  (3.2,-2.0) -- (3,0);
\draw[thin, -latex]  (3.4,-4)--(3.15,-1.5) ;
\draw[very thick, -latex]  (-3,0)--(0,0) ;
\draw[very thick ]  (-0.1,0)--(3,0) ;
\draw(-4,1.5)node[above]{$p_3$};
\draw(-4,-1.5)node[below]{$p_1$};
\draw(4,1.5)node[above]{$p_4$};
\draw(4,-1.5)node[below]{$p_2$};
\draw(0,0.3)node[above]{ $q$};
\draw(0,-0.3)node[below]{ graviton};
\draw(-3.4,-4)node[below]{ $X_J$};
\draw(-3.4,4)node[above]{ $X_J$};
\draw(3.4,-4)node[below]{ $\psi$};
\draw(3.4,4)node[above]{ $\psi$};
\end{scope}
\end{tikzpicture}
\end{center}
\caption{ \small Eikonal scattering of hadrons ($X_J$) in large $N$ gauge theory with a spectating scalar $\psi$. The scalar can only interact with other particles via gravity.}\label{spec_scalar}
\end{figure}
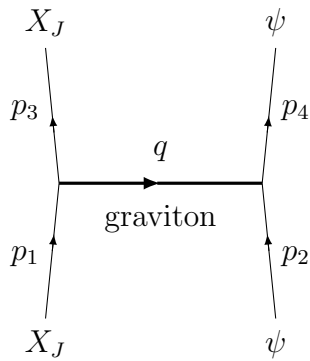

Again we can consider an eikonal scattering of a glueball or meson (which we will denote as $X_J$) of arbitrary spin with the spectator scalar (see figure \ref{spec_scalar}). The phase-shift, even when $N$ obeys  the bound (\ref{bound1}), is completely fixed by the on-shell three-point amplitude $X_JX_J h_{\mu\nu}$. A priori, the on-shell three-point amplitude $X_J X_J h_{\mu \nu}$ can be any linear combination of $2J+1$ parity even and $2J$ parity odd structures (see section 2.5 of \cite{Afkhami-Jeddi:2018apj})
\begin{align}
\langle X_J(p_1,z_1)X_J(p_3,z_3)h(q,z)\rangle&=A^2 \sum_{i=1}^{J+1}a_i (z_1\cdot z_3)^{J-i+1}(z_1\cdot q)^{i-1}(z_3\cdot q)^{i-1}\\
+AB &\sum_{i=1}^{J}a_{J+i+1} (z_1\cdot z_3)^{J-i}(z_1\cdot q)^{i-1}(z_3\cdot q)^{i-1}+\text{parity odd}\ ,\nonumber
\end{align}
where, $a_i$ with $i=1,\cdots, 2J+1$ are the coupling constants and $A=(z\cdot p_3)$, $B=(z\cdot z_3)(z_1\cdot q)-(z\cdot z_1)(z_3\cdot q)$. Positivity of the phase shift for this eikonal scattering strongly constrains the on-shell three-point amplitude $X_JX_J h_{\mu\nu}$. Following section 2.5 of \cite{Afkhami-Jeddi:2018apj}, we conclude that  the parity odd part the on-shell three-point amplitude $X_JX_J h_{\mu\nu}$ must vanish and the parity even part is completely fixed
\begin{align}\label{bounds4}
&\frac{a_{n+1}}{a_n}=\frac{(n-J)(n+J-1)}{n(2n-1)}\frac{1}{m^2}\ , \qquad n=1,\cdots,J\ , \nonumber\\
&\frac{a_{J+n+2}}{a_{J+n+1}}=\frac{n^2-J^2}{n(2n+1)}\frac{1}{m^2}\ , \qquad n=1,\cdots,J-1\ ,
\end{align}
with $a_{J+2}=J a_1$ and $m$ being the mass of $X_J$. Note that the coefficient $a_1$ is also fixed by the soft theorem. Therefore, spinning hadrons of any confining large $N$ gauge theory in $(3+1)$-dimensions must couple to graviton in a specific non-minimal way if we want to include spectator fields such as dark matter. Moreover, in the presence of a spectator field, the mixing bound (\ref{bound_mixing}) also applies to any confining large $N$ gauge theory.

\subsubsection*{Higher Dimensions}
Main results of this paper are derived for confining gauge theories in $(3+1)$-dimensions, however, the analysis can be easily extended to higher dimensions. In fact, causality leads to stronger constraints in higher dimensions. For example, a simple extension of the flat space eikonal scattering argument of \cite{Afkhami-Jeddi:2018apj} (for the setup shown in figure \ref{spec_scalar}) implies that in the presence of a spectator field any confining large $N$ gauge theory would violate causality in $d\ge 5$ spacetime dimensions. This is consistent with the fact that there is no confining gauge theory in $d\ge 5$ dimensions. On the other hand, the entropic argument of section \ref{sec:SpeciesBoundEntropy} suggests that there is a bound on $N$ even for non-confining gauge theories in $d\ge 5$ dimensions.


\section*{Acknowledgements}
It is our pleasure to thank Nima Afkhami-Jeddi, Liam Fitzpatrick, Tom Hartman, Ami Katz, and Amirhossein Tajdini  for several helpful discussions as well as comments on a draft. We would also like to thank  Nima Arkani-Hamed, Ibou Bah, and David Kaplan for discussions. We were supported in part by the Simons Collaboration Grant on the Non-Perturbative Bootstrap.

\appendix
\section{Smearing of Regge Correlator}\label{smearing}
Consider the correlator 
\be
G=\langle \psi(x_1) \O(x_2)\O(x_3) \psi(x_4)\rangle \nonumber
\ee
in $d$-spacetime dimensions, where we choose the points as follows:
\begin{align}\label{points}
&x_1=(u,v,\vec{0})\ , \qquad x_2=(t=i B,y^1=-1,\vec{0})\ , \nonumber\\
&x_4=(-u,-v,\vec{0})\ , \qquad x_3=(t=i(B+\tau),y^1=1,\vec{y})\ ,
\end{align}
First, we take the Regge limit: 
\be
u=\frac{1}{\sigma}\ , \qquad v=-\sigma B^2 \rho \qquad \text{with} \qquad \sigma \rightarrow 0 \ .
\ee
with $0< \rho <1$ and $ B$ fixed. Next, we integrate over $\tau$ and $\vec{y}$ after taking the large $B$ limit:
\be
G_{smeared}= \int_{-\infty}^{\infty} d\tau \int d^{d-2}\vec{y} \lim_{B\rightarrow \infty} \lim_{\sigma\rightarrow 0} G\ .
\ee

\section{Double Trace Operators}\label{app_dt}
We consider two free scalar fields $\phi_1$ and $\phi_2$ in AdS which are dual to two primary scalar operators $\O_1$ and $\O_2$ with dimensions $\Delta$. In this appendix we explicitly construct double trace primary operators 
\be\nonumber
[\O_1 \O_2]_{n,\ell}\sim   \O_1\Box^n \partial_{\mu_1}\partial_{\mu_2}\cdots \partial_{\mu_\ell} \O_2+\cdots
\ee
that are dual to free two-particle states bound only by the effect of the AdS curvature.

\subsection*{$\ell=0$}
Let us first write $[\O_1 \O_2]_{0,0}$
\be
[\O_1 \O_2]_{0,0}=\lim_{x_1\rightarrow x_2} \O_1(x_1)\O_2(x_2) \ .
\ee
\subsection*{$\ell=1$}
For $\ell=1$, let us first introduce the following notation
\be
D_1= (\varepsilon.\partial_2-\varepsilon.\partial_1)\ , \qquad D_0^2= \varepsilon.\partial_1 \varepsilon.\partial_2\ ,
\ee
where, $\varepsilon^\mu$ is the null polarization vector. In this notation, $[\O_1 \O_2]_{0,1}$ is given by
\be
[\O_1 \O_2]_{0,1}=\frac{1}{\sqrt{4\Delta}} \lim_{x_1\rightarrow x_2}  D_1 \O_1(x_1)\O_2(x_2)\ .
\ee

\subsection*{$\ell=2$}
For $\ell=2$, we have
\be
[\O_1 \O_2]_{0,2}=N_2 \lim_{x_1\rightarrow x_2}  \left(D_1^2 -\frac{2}{\Delta} D_0^2\right)\O_1(x_1)\O_2(x_2)\ 
\ee
with
\be
N^2_2=\frac{1}{16 (\Delta +1) (2 \Delta +1)}\ .
\ee

\subsection*{$\ell=3$}
Let us now write $[\O_1 \O_2]_{0,3}$ as follows
\begin{align}
[\O_1 \O_2]_{0,3}=&\lim_{x_1\rightarrow x_2} a_1 \O_1(x_1)(\varepsilon.\partial_2)^3 \O_2(x_2) +a_2 (\varepsilon.\partial_1)\O_1(x_1)(\varepsilon.\partial_2)^2 \O_2(x_2)\nonumber \\
&+a_3 (\varepsilon.\partial_1)^2\O_1(x_1)(\varepsilon.\partial_2) \O_2(x_2)+a_4 (\varepsilon.\partial_1)^3\O_1(x_1) \O_2(x_2)
\end{align}
where, $\varepsilon$ is a null polarization vector. The operator $[\O_1 \O_2]_{0,3}$ has dimension $2\Delta+\ell$ and spin $\ell= 3$. We will determine $a_i$'s by demanding that this operator is primary and the two-point function of this operator is appropriately normalized. In particular, we obtain
\begin{align}
a_2=-a_3=-\left(3+\frac{6}{\Delta}\right)a_1\ , \qquad a_1^2=\frac{1}{192(\Delta +1) (\Delta +2) (2 \Delta +3)}\ ,\qquad a_4=-a_1\ 
\end{align}
Therefore,
\be\label{spin3}
[\O_1 \O_2]_{0,3}=N_3 \lim_{x_1\rightarrow x_2}  \left(D_1^3 -\frac{6}{\Delta} D_0^2 D_1\right)\O_1(x_1)\O_2(x_2)\ 
\ee
with $N_3=a_1$.

\subsection*{$\ell=4$}
Similarly, for $\ell=4$, we have
\be
[\O_1 \O_2]_{0,4}=N_4 \lim_{x_1\rightarrow x_2}  \left(D_1^4 -\frac{12}{\Delta} D_0^2 D_1^2+\frac{12}{\Delta  (\Delta +1)}D_0^4\right)\O_1(x_1)\O_2(x_2)\ 
\ee
with
\be
N_4^2=\frac{1}{1536 (\Delta +2) (\Delta +3) (2 \Delta +3) (2 \Delta +5)}\ .
\ee

\section{Heavy-Heavy-Light-Light Regge Correlator}\label{app_hhll}
In this appendix, we show the derivation of the four-point function $\langle \psi(x_1) \O(x_2)\O(x_3) \psi(x_4)\rangle$ in the Regge limit for holographic CFTs. The scalar primary $\psi$ is heavy: $c_T\gg \Delta_{\psi}\gg 1$, whereas, the operator $\O$ is light. Conformal invariance guarantees that the four-point function can be written in the following form
\be
\langle \psi(x_1) \O(x_2)\O(x_3) \psi(x_4)\rangle=\frac{1}{x_{14}^{2\Delta_\psi}x_{23}^{2\Delta_\O}}G(z,\bz)
\ee
where, cross-ratios $z$ and $\bz$ are defined in the usual way
\begin{align}
z\bz=\frac{x_{12}^2 x_{34}^2}{x_{13}^2x_{24}^2}\ , \qquad (1-z)(1-\bz)=\frac{x_{14}^2x_{23}^2}{x_{13}^2x_{24}^2}\ ,
\end{align}
where, $x_{ij}=|x_i-x_j|^2$. Let us now choose the points as follows:
\begin{align}
&x_1=(u,v,\vec{0})\ , \qquad x_2=(t=0,y^1=-1,\vec{0})\ , \nonumber\\
&x_4=(-u,-v,\vec{0})\ , \qquad x_3=(t=0,y^1=1,\vec{0})\ ,
\end{align}
with 
\be
u=\frac{1}{\sigma}\ , \qquad v=-\sigma \eta  \ .
\ee
In the Regge limit $\sigma \rightarrow 0$ cross-ratios are given by
\be\label{cr7}
\bz=4\eta \sigma\ , \qquad z=4\sigma\ .
\ee
We can calculate the Regge four-point function by using the Regge OPE of $\psi \psi$ following \cite{Afkhami-Jeddi:2017rmx}. For holographic CFTs, at the leading order in $1/c_T$ we obtain 
\be
\frac{\langle \psi(x_1) \O(x_2)\O(x_3) \psi(x_4)\rangle}{\langle \psi(x_1) \psi(x_4)\rangle}= \frac{1}{2^{2\Delta_\psi}}+ \frac{40 \Delta_\O}{c_T \pi^2 \sigma} \int_{-\infty}^{\infty} du \Pi_{uu}(x_1,x_2;u,v=0,\vec{y}=\vec{0},z=\sqrt{\eta})
\ee
where, $ \Pi_{\alpha' \beta'}$ is given by 
\be
\Pi_{\alpha' \beta'}(x_1,x_2;z',x')=-\int d^{d+1}x \sqrt{g^{AdS}} G^{\mu\nu}_{\alpha' \beta'}(z,x; z',x')T^{bulk}_{\mu\nu}(D^\phi(z,x;x_1);D^\phi(z,x;x_2))\ .
\ee
$G^{\mu\nu}_{\alpha' \beta'}(z,x; z',x')$ is the bulk-to-bulk graviton propagator and $T^{bulk}_{\mu\nu}$ is the bulk stress tensor of a scalar field $\phi$ which is dual to the operator $\O$. For, $d=4$, following \cite{DHoker:1999mqo} (see also \cite{Fitzpatrick:2011hh} and \cite{Afkhami-Jeddi:2017rmx}), we can derive
\be\label{pi}
\Pi_{\alpha' \beta'}(x_1,x_2;z',x')=\frac{ \Delta_\O}{4\pi^2}\frac{1}{x_{12}^{2\Delta}}\frac{1}{z'^2}\left(\frac{1}{3}\eta_{\alpha' \beta'}-J_{\alpha' z'}(X'-X_1)J_{\beta' z'}(X'-X_1) \right)f(t)+\cdots\ ,
\ee
where dots represent terms that do not contribute to the final correlator because they are gauge dependent. In the above expression $X'=(z',x')$, $X_1=(z=0,x_1)$ and the inversion tensor is\footnote{Note that indices raised and lowered with $\eta_{\alpha\beta}$.}
\be
J_{\alpha\beta}(X'-X_1)=\eta_{\alpha\beta}-\frac{2(X'-X_1)_{\alpha}(X'-X_1)_{\beta}}{z'^2+|x'-x|^2}\ .
\ee
The function $f(t)$ is given by
\be
f(t)=\frac{t(1-t^{\Delta_\O-1})}{(1-t)}\ , \qquad t=\frac{z'^2 x_{12}^2}{(z'^2+(x'-x_1)^2) (z'^2+(x'-x_2)^2)}\ .
\ee
We can perform the remaining $u$-integral by a residue and for integer $\Delta_\O$, we obtain
\be\label{}
\frac{\langle \psi(x_1) \O(x_2)\O(x_3) \psi(x_4)\rangle}{\langle \psi(x_1)   \psi(x_4)\rangle\langle\O(x_2)\O(x_3) \rangle}= 1- i\left(\frac{10 \Delta_\psi \Delta_\O}{c_T \pi^3}\right)\frac{ \eta P_{2\Delta_\O-4}(\eta)}{\sigma (1+\eta)^{2\Delta_\O-1}}\ ,
\ee
where, $ P_{2\Delta_\O-4}(\eta)$ is a polynomial of degree $(2\Delta_\O-4)$ and $P_{2\Delta_\O-4}(0)=1$. For simplicity, let us restrict to $\Delta_\O=2$ for which $P_0(\eta)=1$. So, at the leading order in $1/c_T$ we find that the Regge correlator is given by
\be\label{regge_corr}
\frac{\langle \psi(x_1) \O(x_2)\O(x_3) \psi(x_4)\rangle}{\langle \psi(x_1)   \psi(x_4)\rangle\langle\O(x_2)\O(x_3) \rangle}\approx 1-i \frac{80 \Delta_\psi}{c_T \pi^3} \frac{z \bz}{(z+\bz)^3}\ ,
\ee
where, we have used equation (\ref{cr7}) to write the final result in terms of conformal cross-ratios. 

\section{Bounding the On-shell Mixing Amplitudes}\label{mixing}
\begin{figure}
\begin{center}
\usetikzlibrary{decorations.markings}    
\usetikzlibrary{decorations.markings}    
\begin{tikzpicture}[baseline=-3pt,scale=0.75]
\begin{scope}[very thick,shift={(4,0)}]
\draw[very thick,-latex]  (0,0) -- (-1.1,1.1);
\draw[very thick]  (-1,1) -- (-2,2);
\draw[very thick]  (-1.1,-1.1) -- (0,0);
\draw[very thick, -latex]  (-2,-2) -- (-1,-1);
\draw[thin,-latex]  (1.7,0.3) -- (2.3,0.3);
\draw  (0,0) circle (1.5pt) ;
\draw [domain=0:3.5, samples=500]  plot (\x, {0.1* sin(5*pi*\x r)});
\draw(-2,2)node[above]{$G'_{J'}/\pi'_{J'}$};
\draw(-1,1)node[right]{$p_3, z_3$};
\draw(-2,-2)node[below]{$G_J/\pi_J$};
\draw(-1,-1)node[right]{$p_1, z_1$};
\draw(2,0.3)node[above]{ $h_{\mu\nu}$};
\draw(3.6,0.0)node[right]{ $q, z$};
\end{scope}
\end{tikzpicture}
\end{center}
\caption{\label{fig_tpf_app} \small The three-point interaction between a glueball/meson with spin $J$, another glueball/meson with spin $J'$ and a graviton.}
\end{figure}
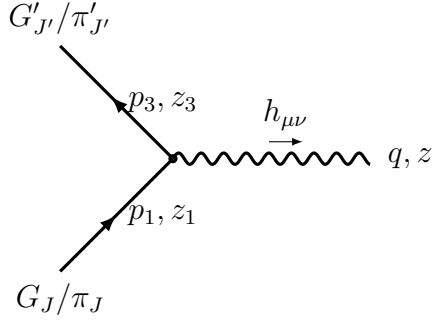

Let us now consider the three-point interaction between a glueball with spin $J$ and mass $m_1$, another glueball with spin $J'$ and mass $m_3$ and a graviton. The on-shell three-point amplitude is completely fixed by symmetries.\footnote{See section 2 of \cite{Afkhami-Jeddi:2018apj} for a review.} The conservation of momentum, on-shell conditions and gauge invariance of the graviton imply that the on-shell three-point amplitude should be constructed using the following building blocks (see figure \ref{fig_tpf_app} for our notations):
\ba\label{eq:bilblock}
&z_1\cdot z_3\ ,  \qquad z_1\cdot q\ , \qquad z_3\cdot q\ , \qquad (z\cdot z_3)(z_1 \cdot q) - (z\cdot z_1)(z_3\cdot q)\equiv B\\
&(z \cdot p_3)(z_3 \cdot q)-\frac{1}{2}(z\cdot z_3)(m_3^2-m_1^2)\equiv A_1\ , \quad  (z \cdot p_3)(z_1 \cdot q)-\frac{1}{2}(z\cdot z_1)(m_3^2-m_1^2)\equiv A_2\ .\nonumber
\ea
Note that in $(3+1)$-dimensions $A_1, A_2$ and $B$ are not completely independent structures. They obey the following identities:
\ba
& (z_1\cdot q)A_1-(z_3\cdot q)A_2=\frac{1}{2}(m_3^2-m_1^2)B\ , \nonumber\\
& m_3^2 B^2+2 (z_1 \cdot z_3) A_1 A_2+2 (z_1\cdot q) A_1 B= 0\ .
\ea

\subsection*{$J'>J$:}
\begin{figure}
\begin{center}
\usetikzlibrary{decorations.markings}    
\usetikzlibrary{decorations.markings}    
\begin{tikzpicture}[baseline=-3pt,scale=0.40]
\begin{scope}[very thick,shift={(4,0)}]
\draw[thin,-latex]  (-3,0) -- (-3.2,2.0);
\draw[thin]  (-3.15,1.5) -- (-3.4,4.0);
\draw[thin]  (-3.2,-2.0) -- (-3,0);
\draw[thin, -latex]  (-3.4,-4)--(-3.15,-1.5) ;
\draw[thin, -latex]  (3,0) -- (3.2,2.0);
\draw[thin]  (3.15,1.5) -- (3.4,4.0);
\draw[thin]  (3.2,-2.0) -- (3,0);
\draw[thin, -latex]  (3.4,-4)--(3.15,-1.5) ;
\draw[very thick, -latex]  (-3,0)--(0,0) ;
\draw[very thick ]  (-0.1,0)--(3,0) ;
\draw(-4,1.5)node[above]{$p_3$};
\draw(-4,-1.5)node[below]{$p_1$};
\draw(4,1.5)node[above]{$p_4$};
\draw(4,-1.5)node[below]{$p_2$};
\draw(0,0.3)node[above]{ $q$};
\draw(0,-0.3)node[below]{ graviton};
\draw(-3.4,-4)node[below]{ $G_J$};
\draw(-3.4,4)node[above]{ $ G'_{J'}$};
\draw(3.4,-4)node[below]{ $G_0$};
\draw(3.4,4)node[above]{ $G_0$};
\end{scope}
\end{tikzpicture}
\end{center}
\caption{\label{app_mix} \small The phase-shift $\delta_{GG'}$ for glueball mixing in the limit $N\rightarrow \infty$ is obtained from this process. }
\end{figure}
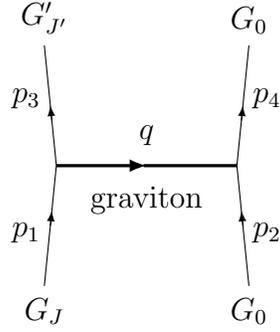

Now the most general form of on-shell amplitude can be represented as a linear combination of the following structures ($J'> J\ge 2$):
 \ba\label{set1}
&\mathcal{A}_1  = A_1^2 (z_1 \cdot z_3)^{J_0}(z_1\cdot q)^{J-J_0} (z_3\cdot q)^{J'-J_0-2}\ , \\
&\mathcal{A}_2 =  A_1^2 (z_1 \cdot z_3)^{J_0-1} (z_1\cdot q)^{J-J_0+1} (z_3\cdot q)^{J'-J_0-1}\ , \nonumber\\
&\vdots \nonumber \\
&\mathcal{A}_{J_0+1} =  A_1^2   (z_1 \cdot q)^J (z_3\cdot q)^{J'-2}\ , \nonumber\\
&\mathcal{A}_{J_0+2} = A_1 B (z_1 \cdot z_3)^{J_0'-1} (z_1\cdot q)^{J-J_0'} (z_3\cdot q)^{J'-J_0'-1}\ , \nonumber\\
&\mathcal{A}_{J_0+3} =  A_1 B (z_1\cdot z_3)^{J_0-2}  (z_1 \cdot q)^{J-J_0'+1}(z_3\cdot q)^{J'-J_0}\ , \nonumber\\
&\vdots \nonumber\\
& \mathcal{A}_{J_0+J_0'+1} = A_1 B  (z_1 \cdot q)^{J_0-1}(z_3\cdot q)^{J'-2}\ .\nonumber
\ea
where, 
\ba
&J_0=min(J,J'-2)\ , \qquad J_0'=min(J,J'-1) \qquad \text{for} \qquad m_1\neq m_3\ , \nonumber\\
&J_0=J_0'=J \qquad \text{for} \qquad m_1=m_3\  .
\ea
Therefore, the three-point amplitude for $J'> J$ is given by
\ba
C_{JJ'2}=  \sqrt{32 \pi G_N} \sum_{n= 1}^{J_0+J_0'+1} a_n \mathcal{A}_n \ ,
\ea 
where, $a_n$'s are coupling constants. This allows us to compute the phase shift $\delta_{GG'}$ (or $\delta_{\pi\pi'}$ for mesons) for the process \ref{app_mix}.
In the limit $\Lambda_{\text{UV}}\gg 1/b\gg m_G, m_{G'}$, the phase-shift grows as 
\be
\delta_{GG'} \sim a_1\frac{s}{M_{Pl}^2} \frac{1}{b^n}\ ,
\ee
where $n\ge 1$. This phase shift violates the interference bound (\ref{intf_bound}) implying that the on-shell mixing amplitudes must be suppressed by the cut-off scale
\be
G'_{J'} G_J h_{\mu \nu} \lesssim \frac{1}{M_{Pl}} \frac{\ln(\Lambda_{\text{UV}} L)}{\Lambda_{\text{UV}}^{n}}\ .
\ee

\subsection*{$J'=J$}
For this case, the set (\ref{set1}) still contributes, however, there is an additional structure which is independent when $m_1\neq m_3$
\be
 \mathcal{B}=A_1 A_2 (z_1 \cdot z_3)^{J-1}\ .
\ee
The most general on-shell three-point amplitude for $J'=J$ is given by
\ba
C_{JJ2}=  \sqrt{32 \pi G_N}\left( \sum_{n= 1}^{2J-2} \tilde{a}_n \mathcal{A}_n + \tilde{b} \mathcal{B} \right)\ .
\ea 
In the limit $\Lambda_{\text{UV}}\gg 1/b\gg m_G, m_{G'}$, the phase-shift now grows at least as fast as 
\be
\delta_{GG'} \sim  \tilde{b}\frac{s}{M_{Pl}^2} \frac{1}{b^2}\ .
\ee
This phase shift again violates the interference bound (\ref{intf_bound}) implying 
\be
G'_{J} G_J h_{\mu \nu} \lesssim \frac{1}{M_{Pl}} \frac{\ln(\Lambda_{\text{UV}} L)}{\Lambda_{\text{UV}}^2}\ .
\ee
For the special case, $J'=J$ and $m_1=m_3$, there is a particular on-shell three-point interaction which is consistent with the growth bound. However, this interaction does not have the right soft limit for non-identical particles. In particular, if we impose that the amplitude $G'_{J'} G_J h_{\mu \nu}$ has the right soft limit: $G'_{J} G_J h_{\mu \nu}(q)\rightarrow 0$ when $q\rightarrow 0$ -- that necessarily requires a $\Lambda_{\text{UV}}$ suppression 
\be
G'_{J} G_J h_{\mu \nu} \lesssim \frac{1}{M_{Pl}} \frac{\ln(\Lambda_{\text{UV}} L)}{\Lambda_{\text{UV}}} \ .
\ee
Note that the analysis of this appendix holds for meson mixing $\pi\pi' h_{\mu\nu}$ as well.

\end{spacing}

\bibliographystyle{utphys} 
\bibliography{CausalityBib}

\end{document}